\documentclass[footinbib,a4paper,aps,prA,reprint,twocolumn,preprintnumbers,amsmath,amssymb,10pt, superscriptaddress]{revtex4-1}
\usepackage{physics}
\usepackage{verbatim}
\usepackage{graphicx,wrapfig,lipsum}
\usepackage{color}
\usepackage{tikz}
\usepackage[colorlinks=true,citecolor=blue,linkcolor=blue,urlcolor=blue]{hyperref}
\usepackage{relsize}
\usepackage{braket}
\usepackage[T1]{fontenc}
\usepackage[normalem]{ulem}
\usepackage{upgreek}
\usepackage[percent]{overpic}
\usepackage{graphicx,xcolor}
\usepackage{dcolumn}
\usepackage{bm}
\usepackage[shortlabels]{enumitem}
\graphicspath{{Figs/}}
\makeatletter
\renewcommand{\fnum@figure}{FIG. \thefigure}
\makeatother

\begin{document}

\title{Emergent superconformal symmetry \\ in the phase diagram of a 1D \texorpdfstring{$\mathbb{Z}_{2}$}{TEXT} lattice gauge theory}
\author{Bachana Beradze}
\affiliation{Department of Engineering and Physics, Karlstad University, Karlstad, Sweden}
\affiliation{Andronikashvili Institute of Physics, Tamarashvili str. 6, 0177 Tbilisi, Georgia}
\affiliation{Ilia State University, Cholokashvili Avenue 3-5, 0162 Tbilisi, Georgia}
\email{bachana.beradze@kau.se}

\author{Mikheil Tsitsishvili}%
\affiliation{Institut f\"ur Theoretische Physik, Heinrich-Heine-Universit\"at, D-40225  D\"usseldorf, Germany}
\email{mikheil.tsitsishvili@hhu.de}

\author{Sergej Moroz}
\affiliation{Department of Engineering and Physics, Karlstad University, Karlstad, Sweden}
\affiliation{Nordita, KTH Royal Institute of Technology and Stockholm University, Stockholm, Sweden}
\email{sergej.moroz@kau.se}

\date{\today}

\begin{abstract}
    We investigate the phase diagram and critical properties of a one-dimensional $\mathbb{Z}_{2}$ lattice gauge theory describing an orthogonal metal, where spinless fermions and Ising spins are minimally coupled to a deconfined $\mathbb{Z}_{2}$ gauge field.
    Working at half-filling of fermions, we derive an exact gauge-invariant formulation that maps the model onto decoupled XXZ and transverse-field Ising chains.
    This mapping enables a controlled low-energy field-theory description in terms of a perturbed Luttinger liquid and Ising conformal field theories.
    Combining analytical arguments with numerical simulations, we determine the full phase diagram and identify various critical and multi-critical regimes.
    Along a specific multi-critical line, where the fermionic and bosonic velocities coincide, we find strong evidence for an emergent superconformal symmetry.
    Our results establish a minimal lattice realization of emergent superconformal criticalities in a gauge-matter system and provide a route toward its exploration in quantum simulators.
\end{abstract}

\maketitle
\section{Introduction}
\label{sec:introduction}

Recent breakthroughs in quantum simulation with ultracold atoms in optical lattices, trapped ions, superconducting circuits, and programmable Rydberg-atom arrays have opened a new chapter in the study of quantum lattice gauge theories (LGTs), where local constraints and gauge--matter interactions can be engineered and probed directly in real time.
Such out-of-equilibrium settings are beyond the reach of many classical methods.
These developments gave impetus to an intense effort to implement LGTs with dynamical matter and to access hallmark phenomena such as confinement, string breaking, and topological order in controllable table-top experiments, see Refs.~\cite{Zohar2016Review,Wiese2013Review,BanulsCichy2020Review,Halimeh2025Review} for general overviews.
Particularly prominent in this context are $\mathbb{Z}_{2}$ gauge theories: they admit minimal local Hilbert spaces while retaining genuinely gauge-invariant dynamics, and they naturally arise in a variety of platforms, enabling experimental and Floquet-engineered realizations, as well as resource-efficient digital protocols for $\mathbb{Z}_{2}$ gauge fields coupled to matter~\cite{Schweizer2019NaturePhys,Zohar2017PRL, Bazavan2023arXiv, Homeier2023CommsPhys, ConfinementZ2QuantumComputer2025NatPhys}.

Motivated by these developments, we focus on a minimal gauge--matter setting that realizes an orthogonal metal~\cite{NandkishoreMetlitskiSenthil2012PRB}.
The key idea is a $\mathbb{Z}_{2}$ fractionalization of the physical fermion creation operator,
$c^{\dagger}=f^{\dagger}\tau^{z}$,
into an ``orthogonal'' fermion $f^{\dagger}$ and an Ising spin-$\flatfrac{1}{2}$ operator $\tau^{z}$. 
This decomposition has a discrete $\mathbb{Z}_{2}$ redundancy because one can multiply each constituent by a minus sign without changing $c^{\dagger}$.
Promoting this redundancy to a local constraint yields a $\mathbb{Z}_{2}$ lattice gauge theory, where both $f$ and $\tau$ carry $\mathbb{Z}_{2}$ gauge charge and are minimally coupled to an Ising gauge field~\cite{SenthilFisher2000Confinement,FradkinShenker1979}.
This setting provides a direct route to metallic phases whose low-energy fermionic excitations are orthogonal to the gauge-invariant quasiparticles created by $c^{\dagger}$.

In this paper, we investigate quantum phases of a one-dimensional realization of the orthogonal metal problem.
To this end, we first exactly resolve the Gauss-law constraint of the microscopic model, which makes numerical simulations of the original lattice model straightforward.
Moreover, we establish a non-local Jordan-Wigner transformation that maps the microscopic Hamiltonian to a direct sum of spin--$\flatfrac{1}{2}$ XXZ and quantum Ising chains.
Both of these paradigmatic one-dimensional models have been extensively studied using conformal field theory (CFT)~\cite{Belavin1984,Senechal1997,Ginsparg1988}, bosonization~\cite{Giamarchi2003,Nersesyan2004}, and state-of-the-art tensor-network methods~\cite{White1992,Schollwock2011,Orus2014,Verstraete2008}.
As a result, the exact non-local mapping provides analytic control over locations and universality classes of phase transition critical lines.

Beyond the characterization of phases and phase transitions in the orthogonal metal model, we investigate the emergence of enhanced symmetries at special points in the phase diagram.
The emergent symmetries in the low-energy sectors of microscopic models play a central role in modern condensed-matter and high-energy physics.
Of particular interest is the emergence of supersymmetry (SUSY), where the bosonic and fermionic sectors of the theory transform onto each other.
The lattice realizations of
SUSY have been identified in several contexts, including interacting fermion chains and quantum spin systems~\cite{Fendley2003-1,Fendley2003-2,OBrien_Fendley_2018-3,Rahmani2015,Dalmonte2015,Hsieh2016,Tsitsishvili2022,Fromholz2022,Wu2025}.
A paradigmatic example is the multi-critical point ~\cite{Bauer2013,Alberton2017,Huijse2015} where a massless compact bosonic theory and Ising criticalities coexist, which is described by a $\mathcal{N}=(1,1)$ superconformal field theory.
A more exotic scenario occurs when a Berezinskii–Kosterlitz–Thouless (BKT) transition and critical Ising theory meet, which is described by a $\mathcal{N}=(3,3)$ superconformal field theory.

In this work, we demonstrate that the one-dimensional $\mathbb{Z}_{2}$ lattice gauge theory realizing an orthogonal metal hosts a critical line with emergent ${\mathcal{N}=(1,1)}$ superconformal symmetry~\cite{Friedan1984,Bershadsky1985,Dixon_Ginsparg_1988}.
The two endpoints of the critical line correspond to ${\mathcal{N}=(3,3)}$ SUSY~\cite{Schwimmer1987,Ozer2025,Chang1987} and SU(2)$_{2}$ WZNW universality classes~\cite{Witten1984,Nersesyan2004,Senechal1997}.
Due to the exact mapping to decoupled XXZ and Ising chains, the microscopic gauge-invariant degrees of freedom are related non-locally to the fields of the supersymmetric CFT, providing a concrete lattice realization of the emergent SUSY in a gauge-theory setting.
The theoretical predictions are corroborated by numerical DMRG simulations by extracting the central charges of the underlying CFT and the corresponding fermionic and bosonic excitation velocities.

The paper is organized as follows.
In Sec.~\ref{sec:system} we introduce the model, derive its gauge-invariant formulation, and discuss its symmetries.
In Sec.~\ref{sec:analytics} we present the exact non-local mapping that decouples the theory into the XXZ and Ising sectors and develop the corresponding low-energy field theory.
In Sec.~\ref{sec:phase_diagram} we present the resulting phase diagram and analyze the nature of the phase transitions.
Section~\ref{sec:numerics} contains numerical DMRG results that support analytical predictions within the gapped and gapless phases, as well as at the ${\mathcal{N}=(3,3)}$ SUSY and SU(2)$_{2}$ WZNW supersymmetric critical points. We draw our conclusions in Sec.~\ref{sec:conclusion}.

\section{The System}
\label{sec:system}
\subsection{The model}
We study a one-dimensional orthogonal metal, a model of interacting one-component fermions and spin-$\flatfrac{1}{2}$ degrees of freedom that are coupled to $\mathbb{Z}_{2}$ gauge field.
It is defined by the following Hamiltonian:
\begin{equation}
    \label{eq:Orthogonal-Metal}
    \begin{split}
        H = &-t \sum_{j}
        \left(
            f^{\dag}_{j}
            \sigma^{z} _{j+\frac{1}{2}}
            f^{\phantom{\dag}}_{j+1}
            + \text{h.c.}
        \right)
        \\
        &+V \sum_{j}
        \left(n_{j}-\frac{1}{2} \right)
        \left(n_{j+1}-\frac{1}{2} \right)
        \\
        &-J \sum_{j}
        \tau^{z}_{j} \sigma^{z}_{j+\frac{1}{2}} \tau^{z}_{j+1}
        -\Gamma \sum_{j} \tau^{x}_{j}.
    \end{split}
\end{equation}
Here, $f^{\dag}_{j}$ and $f^{\phantom{\dag}}_{j}$ are the creation and annihilation operators of a spinless fermionic matter field, defined at each site $j$.
The composite $n_{j} = f^{\dag}_{j}f^{\phantom{\dag}}_{j}$ is the fermion occupation number.
Additionally, we have a spin matter field $\tau^{x,y,z}_{j}$ residing at every site.
The coupling between the spin and the fermionic matter fields is mediated by a $\mathbb{Z}_{2}$ gauge degree of freedom $\sigma^{x,y,z}_{j+\flatfrac{1}{2}}$, defined on links $j + \flatfrac{1}{2}$, connecting sites $j$ and $j+1$.
Both $\tau^{\alpha}$ and $\sigma^{\alpha}$ ($\alpha = x,y,z$) are mutually commuting sets of Pauli matrices, obeying a standard SU(2) algebra.
The Hamiltonian, Eq.~\eqref{eq:Orthogonal-Metal}, consists of a fermion and spin parts:
In the fermionic sector, the tunneling amplitude $t$ minimally couples fermions and the gauge field through the Peierls substitution.
Fermions residing at the nearest-neighbor sites interact with the density-density interaction $V$.
The $\tau$-spin sector consists of a $\Gamma$ magnetic field along the $x$-axis and an Ising type exchange interaction $J$, which minimally couples the spins and the gauge field through the Peierls substitution.
In the Hamiltonian Eq.~\eqref{eq:Orthogonal-Metal}, the gauge fields appear only via the minimal coupling, while the dynamical (electric) terms are not included. The gauge sector is thus in the deconfined regime.
We work at a particle-hole (PH) symmetric regime of a half-filled band for fermions, without imposing additional constraints on the spin matter field and the gauge degree of freedom.
The system is thought to be infinitely large unless specified otherwise.

The Hamiltonian in Eq.~\eqref{eq:Orthogonal-Metal} is invariant under a local $\mathbb{Z}_{2}$ gauge transformation generated by
\begin{equation}
\label{eq:Gauge-Generators}
    G_{j} = \sigma^{x}_{j-\frac{1}{2}}(-1)^{n_{j}}\tau^{x}_{j}\sigma^{x}_{j+\frac{1}{2}}.
\end{equation}
These generators act on the matter fields as
\begin{equation}
    \begin{split}
        f_{j} \rightarrow f'_{j}
            &= G_{j}^{-1} f_{j}G_{j}= -f_{j},
        \\
        \tau^{y,z}_{j} \rightarrow \tau'^{y,z}_{j}
            &= G_{j}^{-1} \tau^{y,z}_{j} G_{j}
            = -\tau^{y,z}_{j},
    \end{split}
\end{equation}
while
\begin{equation}
    \sigma^{y,z}_{j\pm\frac{1}{2}}
        \rightarrow
        \sigma'^{y,z}_{j\pm\frac{1}{2}}
        = G_{j}^{-1} \sigma^{y,z}_{j\pm\frac{1}{2}} G_{j}
        = -\sigma^{y,z}_{j\pm\frac{1}{2}}.
\end{equation}
Due to the local gauge invariance, the Hilbert space is split into $2^{L}$ distinct sectors characterized by $G_{j} = \pm 1$, each of which evolves independently.
We will work in the sector without static charges. Throughout the paper, we enforce the strict constraint ${G_{j} = 1}$ at every lattice site, known as the (even) Gauss law.

We solve the Gauss law for $\tau_{j}^{x}$
\begin{equation}
    \label{eq:even_Gauss_law}
    \tau^{x}_{j} = \sigma^{x}_{j-\frac{1}{2}}(-1)^{n_{j}}\sigma^{x}_{j+\frac{1}{2}},
\end{equation}
and express Eq.~\eqref{eq:Orthogonal-Metal} in terms of gauge-invariant degrees of freedom as
\begin{equation}
    \label{eq:H-in-Gauge-Invariant-Fields}
    \begin{split}
        H = &-t \sum_{j}
        \left(
            c^{\dag}_{j}
            Z_{j+\frac{1}{2}}
            c^{\phantom{\dag}}_{j+1}
            + \text{h.c.}
        \right)
        \\
        &+V \sum_{j}
        \left( n_{j}-\frac{1}{2} \right)
        \left( n_{j+1}-\frac{1}{2} \right)
        \\
        &-J \sum_{j} Z_{j+\frac{1}{2}}
        -\Gamma \sum_{j}
        X_{j-\frac{1}{2}} (-1)^{n_{j}} X_{j+\frac{1}{2}},
    \end{split}
\end{equation}
with a new gauge-invariant fermionic field defined on sites
\begin{equation}
    \label{eq:c-fermions}
    c_{j} \equiv f_{j} \tau^{z}_{j},
    \quad
    n_{j} \equiv c^{\dag}_{j} c^{\phantom{\dag}}_{j} = f^{\dag}_{j} f^{\phantom{\dag}}_{j}
\end{equation}
and new gauge-invariant spin-$\flatfrac{1}{2}$ Pauli operators defined on links
\begin{equation}
    \label{eq:XYZ-spins}
    X_{j+\frac{1}{2}} \equiv \sigma^{x}_{j+\frac{1}{2}},
    \quad
    Z_{j+\frac{1}{2}} \equiv \tau^{z}_{j} \sigma^{z}_{j+\frac{1}{2}}\tau^{z}_{j+1}.
\end{equation}
The Hamiltonian in Eq.~\eqref{eq:H-in-Gauge-Invariant-Fields} represents the main subject of study in this paper.
\subsection{Symmetries}
\label{sec:Symmetries}
In a chain with $L$ sites, the total number of fermions ${N=\sum_{j=1}^{L}c^{\dag}_{j}c_{j}}$ is conserved as a result of the global U(1) symmetry ${c_{j}\to e^{\mathrm{i}\alpha}c_{j}}$. In case of periodic boundary conditions, the model becomes translationally invariant.

Throughout the paper, we work at half-filling ${N=L/2}$ ${(L\in2\mathbb{N})}$.
This case is of particular importance due to the appearance of the PH symmetry generated by:
\begin{equation}
    \begin{split}
        C c_{j} C^{-1} &= (-1)^{j} c_{j}^{\dag},
        \\
        C X_{j + \frac{1}{2}} C^{-1}
            &= (-1)^{j} X_{j + \frac{1}{2}}.
    \end{split}
\end{equation}
Note that in our model the PH symmetry acts not only on fermions but on spin degrees of freedom, which is indeed necessary to compensate the sign flip of the $\Gamma$ term in the Hamiltonian Eq.~\eqref{eq:H-in-Gauge-Invariant-Fields}.
The generator $C$ acts in the Hilbert space as a unitary operator:
\begin{align}
    C &= C_{0} W_{0}, \quad C^{2} = C C^{\dag} = 1,
    \\
\label{eq:PH-Ungauged}
    C_{0} &= \mathrm{i}^{L}
        \prod_{j=1}^{L}
        \left(
            c_j^{\phantom{\dag}}+
            (-1)^{j}
            c_j^{\dag}
        \right),
    \quad
    W_{0} = \prod_{j=1}^{L/2} Z_{2j-\frac{1}{2}},
\end{align}
forming a $\mathbb{Z}_{2}$ group, which will be denoted by $\mathbb{Z}_{2}^{C}$.

Additionally, the spin sector enjoys the so-called magnetic $\mathbb{Z}_{2}$ symmetry, generated by:
\begin{equation}
\label{eq:W-Transformation}
    W = \prod_{j=1}^{L}Z_{j + \frac{1}{2}}.
\end{equation}
The $W$-transformation globally flips the sign of $X_{j + \flatfrac{1}{2}}$. We denote magnetic symmetry group by $\mathbb{Z}_{2}^{W}$.

\section{Analytical treatment}
\label{sec:analytics}
\subsection{Decoupling into XXZ and Ising Sectors}
The Hamiltonian in Eq.~\eqref{eq:H-in-Gauge-Invariant-Fields} can be decomposed into a sum of mutually commuting, exactly solvable components: a spin-$\flatfrac{1}{2}$ Heisenberg XXZ chain and quantum Ising (QI) models.
This must not be surprising in light of the model symmetries: the fermion sector has the U(1)$\otimes\mathbb{Z}_{2}^{C}$ symmetry, while the spin sector is $\mathbb{Z}_{2}^{W}$ invariant. Using non-local transformation (see Appendix~\ref{app:XXZ-QI_mapping} for details)
\begin{equation}
\label{eq:nonlocal-mapping}
    \begin{split}
        c^{\phantom{\dag}}_{j}
            = \frac{\mu^{x}_{j}-i\mu^{y}_{j}}{2}
            \prod^{j-1}_{k=1}
            & ( -\mu^{z}_{k} )
            \prod_{k=0}^{j-1}
            ( -\mu^{z}_{k+\frac{1}{2}} ),
        \\
        X_{j+\frac{1}{2}}
            {=}\, \mu^{x}_{j+\frac{1}{2}}
            \prod^{j}_{k=1}
            ( -\mu^{z}_{k} ),
        & \quad
        Y_{j+\frac{1}{2}}
            {=}\, \mu^{y}_{j+\frac{1}{2}}
            \prod^{j}_{k=1}
            ( -\mu^{z}_{k} ),
        \\
        n_{j} = \frac{1}{2} + \frac{1}{2} \mu^{z}_{j},
        & \quad
        Z_{j+\frac{1}{2}} = \mu^{z}_{j+\frac{1}{2}}.
    \end{split}
\end{equation}
we arrive at the following Hamiltonian in the decoupled form:
\begin{equation}
\label{eq:Decoupled-Model}
    H = H_{\text{XXZ}} + H_{\text{QI}},
\end{equation}
where
\begin{equation}
\label{eq:H_XXZ}
    H_{\text{XXZ}}
        = \frac{t}{2} \sum_{j}
        \left(
            \mu^{x}_{j} \mu^{x}_{j+1} +
            \mu^{y}_{j} \mu^{y}_{j+1}
        \right) +
        \\
        \frac{V}{4} \sum_{j} \mu^{z}_{j} \mu^{z}_{j+1}
\end{equation}
corresponds to a Heisenberg spin chain with spins residing on the sites, while
\begin{equation}
    \label{eq:H_QI}
    H_{\text{QI}}
        = \Gamma \sum_{j}
        \mu^{x}_{j-\frac{1}{2}} \mu^{x}_{j+\frac{1}{2}}-
        J \sum_{j} \mu^{z}_{j+\frac{1}{2}},
\end{equation}
describes Ising spins located on the links.

In what follows, we consider only the case of repulsive fermions $V > 0$.
Without loss of generality, we also assume ${t,\Gamma,J\geq0}$, and introduce the following dimensionless parameters:
\begin{equation}
    \Delta \equiv \frac{V}{2t},
    \quad
    g \equiv \frac{\Gamma}{J},
    \quad
    \kappa \equiv \frac{\Gamma}{t},
\end{equation}
where $\Delta$ is the anisotropy parameter of the XXZ model, while $g$ is the dimensionless Ising coupling.
Additionally, as elaborated in Sec.~\ref{sec:Effctive-Model}, $\kappa$ measures the ratio of excitation velocities in the two models.

It follows straightforwardly that the original lattice symmetries are mapped to the following transformations in terms of $\mu$-spins:
\begin{itemize}
    \item the particle number U(1) symmetry becomes spin rotation acting as:
    \begin{equation}
        \mu^{\pm}_{j} \to e^{\mp\mathrm{i} \alpha} \mu^{\pm}_{j};
    \end{equation}
    \item the particle-hole $\mathbb{Z}_{2}^{C}$ symmetry becomes spin inversion acting as:
    \begin{equation}
        \begin{split}
            (\mu^{x}_{j},\mu^{y}_{j},\mu^{z}_{j}) 
                \to
                (-\mu^{x}_{j},\mu^{y}_{j},-\mu^{z}_{j}),
        \end{split}
    \end{equation}
    \item the magnetic $\mathbb{Z}_{2}^{W}$ symmetry acts as:
    \begin{equation}
        W = \prod_{j=1}^{L} \mu^{z}_{j+\flatfrac{1}{2}}.
    \end{equation}
\end{itemize}

\subsection{The effective low energy description}
The paradigmatic models of one-dimensional XXZ and QI chains have been extensively studied using various analytical~\cite{bethe1931,onsager1952,Giamarchi2003,Nersesyan2004,Fradkin2013,Senechal1997} and numerical methods~\cite{tenpy,Chepiga_Mila_2017}.
Both models are exactly solvable~\cite{bethe1931,Giamarchi2003,Nersesyan2004,Fradkin2013}, with their phase diagram, the nature of the phase transitions, and all the correlation functions fully characterized.

The phase diagram of the XXZ model (or equivalently, of the interacting spinless fermion model in Eq.~\eqref{eq:XXZ-in-spinless-fermions}) is well understood~\cite{Giamarchi2003,Nersesyan2004,Fradkin2013}: for $|\Delta|\leq1$, the system remains critical, exhibiting power-law correlations with $\Delta$-dependent exponents, i.e., the Luttinger liquid (LL) phase.
At $\Delta=-1$, the XXZ chain undergoes a first-order transition into a ferromagnetic phase $(\Delta<-1)$ with finite total magnetization.
At $\Delta=1$, on the contrary, the model undergoes a continuous quantum phase transition into a gapped N\'eel phase $(\Delta>1)$ with spontaneously broken $\mathbb{Z}_{2}^{C}$ symmetry.
This symmetry-broken phase hosts long-range charge-density-wave (CDW) order, corresponding to staggered magnetization in the spin representation.
The transition at $\Delta = 1$ belongs to the BKT universality class.
In this work, since we focus on repulsive spinless fermions $V \geq 0$, we restrict our attention to $\Delta \geq 0$.

Similarly, QI exhibits a phase transition upon crossing $g=1$. 
For $g>1$, 
the system develops a doubly-degenerate ferromagnetically ordered phase with nonzero spontaneous magnetization $\langle \mu^{x}_{i} \rangle \neq 0$ oriented along $\pm x$.
For $g<1$, the system is paramagnetic with nonzero magnetization along the $+z$ direction, which is usually called the disordered phase.

\subsubsection{Effective Field Theory}
\label{sec:Effctive-Model}
In this section, we present the low-energy theory of the model in Eq.~\eqref{eq:H-in-Gauge-Invariant-Fields}.
Since it decouples into XXZ and QI parts, so does the corresponding effective action:
\begin{equation}
\label{eq:Effevtive-S}
    \mathcal{S}[\phi,\psi]
        = \mathcal{S}_{\text{B}}[\phi]
        + \mathcal{S}_{\text{F}}[\psi].
\end{equation}

Here $\mathcal{S}_{\text{B}}[\phi]$ describes the infrared limit of the XXZ sector, which in the Euclidean formalism is given by the sine-Gordon model~\cite{Giamarchi2003,Nersesyan2004,Fradkin2013}:
\begin{equation}
\label{eq:XXZ-Action}
    \mathcal{S}_{\text{B}}[\phi]
        = \mathcal{S}_{\text{B}}^{*}[\phi]
        + \frac{v\Delta}{2\pi^{2}}
        \int \dd{\tau} \dd{x}
        {:}\cos{ \sqrt{4K} \phi }{:}
\end{equation}
of a compactified boson field,
\begin{equation}
    \phi \sim \phi + 2\pi R,
\end{equation}
with the compactification radius $R = 1 / \sqrt{K}$ with $K$ being the so-called Luttinger parameter.
The Gaussian contribution reads:
\begin{equation}
\label{eq:Gaussian-Model}
    \mathcal{S}_{\text{B}}^{*}[\phi]
        = \frac{v}{8\pi}
        \int \dd{\tau} \dd{x}
        \bigg[
            \frac{1}{v^{2}}
            ( \partial_{\tau} \phi )^{2}
            + ( \partial_{x} \phi )^{2}
        \bigg],
\end{equation}
while the colon notation in the perturbing cosine term denotes the usual normal ordering.
The action in Eq.~\eqref{eq:XXZ-Action} is controlled by two parameters: the sound velocity $v$ and the Luttinger parameter $K$, which are exactly obtained from the Bethe ansatz solution~\cite{Lukyanov1998,Sirker_2006}:
\begin{align}
\label{eq:LL-Velocity}
    v(\Delta) &= \frac{\pi v_{0}}{2}
        \frac{ \sqrt{1-\Delta^{2}} }{ \arccos{\Delta} },
    \quad
    v_{0} \equiv 2t a_{0},
    \\
\label{eq:LL-Parameter}
    K(\Delta) &= \frac{\pi}{2 (\pi-\arccos{\Delta}) },
    \quad
    |\Delta|\leq1.
\end{align}

The cosine perturbation appearing in $\mathcal{S}_{\text{B}}[\phi]$ carries a scaling dimension $d_{g} = 4K$.
Within the Luttinger liquid regime $|\Delta|<1$ we have  $d_{g} > 2$, implying that the perturbation is RG-irrelevant and may be neglected.
This way, the gapless XXZ sector corresponds to a compact boson CFT with central charge $c_{\text{B}} = 1$ and is described by the fixed point action in Eq.~\eqref{eq:Gaussian-Model}.
On the other hand, for $\Delta > 1$, we get $d_{g}<2$, and thus the cosine perturbation becomes relevant, driving the system to a strong coupling regime with a dynamically generated mass gap and vanishing central charge.
This spontaneously breaks the $\mathbb{Z}_{2}^{C}$ symmetry, developing a long range CDW order and exponentially decaying correlation functions.
The critical point of $\Delta = 1$ belongs to the BKT universality class described by the SU(2)$_{1}$ WZNW theory perturbed by a marginally irrelevant perturbation.
This transition occurs across the solid and dashed red lines on Fig.~\ref{fig:PhaseDiagramGaugeInvariant}, for any finite value of $\kappa$.

The behavior of the QI sector is captured by a Majorana fermion field theory~\cite{Senechal1997, Nersesyan2004}.
The action functional is expressed as a perturbed CFT:
\begin{equation}
\label{eq:QI-Action}
    \mathcal{S}_{\text{F}}[\psi]
        =\mathcal{S}_{\text{F}}^{*}[\psi]
        + \frac{\mathrm{i} m}{2\pi}
        \int \dd{\tau} \dd{x}\,
        \psi \bar{\psi},
\end{equation}
where the fixed-point action
\begin{equation}
\label{eq:Ising-CFT-Action}
    \begin{split}
        \mathcal{S}_{\text{F}}^{*}[\psi]
            = \frac{u}{4\pi}
            \int \dd{\tau} \dd{x}
            \bigg[
                ~&\psi
                \left(
                    \frac{1}{u}\partial_{\tau}
                    + \mathrm{i} \partial_{x}
                \right)
                \psi
                \\
                + &\bar{\psi}
                \left(
                    \frac{1}{u}\partial_{\tau}
                   -\mathrm{i} \partial_{x}
                \right)
                \bar{\psi} ~
            \bigg],
    \end{split}
\end{equation}
represents the QI model at criticality, with $\psi$ and $\bar{\psi}$ denoting left- and right-moving Majorana field components, respectively.
The corresponding excitation velocity and mass are
\begin{equation}
    u \equiv 2\Gamma a_{0},
    \quad
    m \equiv 2J (1-g).
\end{equation}
The mass term is always strongly relevant and gaps the Ising sector of the system, thus for $m\neq 0$ we have vanishing central charge $c_{\text{F}} = 0$.
On the other hand, at the critical point of $m = 0$, the central charge is $c_{\text{F}} = \flatfrac{1}{2}$.
This transition belongs to the Ising universality class and occurs across the solid and dashed blue lines on Fig.~\ref{fig:PhaseDiagramGaugeInvariant}, for any finite $\kappa$.

In the quadrant on the surface above the solid blue line on Fig.~\ref{fig:PhaseDiagramGaugeInvariant}, the QI and XXZ sectors are simultaneously gapless, $0 \leq \Delta \leq 1$ and $m = 0$.
This multi-critical theory has a total central charge of ${c_{\text{tot}} = c_{\text{B}} + c_{\text{F}} = \flatfrac{3}{2}}$.
Most importantly, for
\begin{equation}
\label{eq:SUSY-Condition}
    \kappa = \kappa_{\text{SUSY}}
        = \frac{\pi}{2}\frac{\sqrt{1-\Delta^{2}}}{\arccos{\Delta}},
\end{equation}
the velocities of the fermionic and bosonic sectors coincide (on the dotted blue line).
In this case, the theory exhibits an emergent superconformal symmetry, governed by the following action:
\begin{equation}
\label{eq:SUSY-Action-Functional}
    \begin{aligned}
        \mathcal{S}_{\text{SUSY}} & [\phi,\psi]
            \\
            & = \frac{1}{2\pi}
            \int \dd[2]{z}
            \left(
                \partial_{z} \phi
                \partial_{\bar{z}} \phi +
                \psi \partial_{\bar{z}} \psi
                + \bar{\psi} \partial_{z} \bar{\psi}
            \right).
    \end{aligned}
\end{equation}
Here, $\partial_{z}$ and $\partial_{\bar{z}}$ denote derivatives with respect to the complex coordinates ${z = v \tau + \mathrm{i} x}$ and ${\bar{z} = v \tau - \mathrm{i} x}$ and ${\dd[2]{z} \equiv v\dd{\tau}\dd{x}}$.
Throughout the interval $0 \leq \Delta < 1$, the system possesses $\mathcal{N} = (1,1)$ superconformal symmetry.
At the endpoint $\Delta=0$ (a noninteracting case with $K=1$), the bosonic contribution in ${\mathcal{S}_{\text{SUSY}}[\phi,\psi]}$ can be reformulated in terms of two massless Majorana fermions; consequently, it falls into the SU(2)$_{2}$ WZNW universality class.
The supersymmetric line ends at $\Delta = 1$ (BKT point with $K = \flatfrac{1}{2}$), where the supersymmetry is enhanced to $\mathcal{N} = (3,3)$.

For further discussions on conformal field theories of the compact boson and Majorana fields, the SU(2)$_{k}$ WZNW models, and superconformal theories, see Appendices~\ref{appendix:CFTs} and~\ref{appendix:CFTs-With-Symmetries} and references therein.

\subsubsection{Observables}
The LL and CDW phases can be probed by studying the structure factor
\begin{equation}
    \label{eq:Structure_factor}
    S(q) = \frac{1}{L} \sum_{j,j'} e^{iq(j-j')}
        \left\langle\left\langle \mu^{z}_{j} \mu^{z}_{j'} \right\rangle\right\rangle,
\end{equation}
with $\left\langle\left\langle \mu^{z}_{j} \mu^{z}_{j'} \right\rangle\right\rangle=\left\langle \mu^{z}_{j} \mu^{z}_{j'} \right\rangle-\left\langle \mu^{z}_{j}\rangle \langle \mu^{z}_{j'} \right\rangle$. Within the LL regime and for small momenta $q \ll 2k_{F}$ ($k_{F}$ being the Fermi-momentum), the structure function scales as
\begin{equation}
    \label{eq:Structure_factor_in_LL}
    S(q \ll 2k_{F}) \approx \frac{2K}{\pi} |q|.
\end{equation}
On the other hand, in the CDW phase $S(q)$ should develop a strong peak at momenta that characterizes the underlying long-range order.
In our case, one should observe a peak at $q=2k_{F}$.
For systems of finite size $L$, the amplitude of the peak should scale as $S(2k_{F}) \sim L^{\alpha}$, with $\alpha \to 1$ for the true long-range order \cite{Giamarchi2003,Sachdev2000,Fradkin2013}.

Disordered and ordered Ising phases can be diagnosed via order and disorder parameters, respectively. In particular, we will use  the parity operator $\mathcal{O}_{P}$ and the two-point correlation function of the local order parameter $\mathcal{O}_{xx}$~\cite{Santoro2024}:
\begin{equation}
    \label{eq:O_P,O_xx}
    \begin{split}
        \mathcal{O}_{P,i,r} &= \left\langle \prod^{i+r}_{j=i}\mu^{z}_{j+\frac{1}{2}} \right\rangle, 
        \\
        \mathcal{O}_{xx,i,r} &= \left\langle \mu^{x}_{i+\frac{1}{2}}\mu^{x}_{i+\frac{1}{2}+r} \right\rangle.
    \end{split}
\end{equation}
In the disordered phase one finds $\mathcal{O}_{P,i,r} \neq 0$ and $\mathcal{O}_{xx,i,r} = 0$, whereas in the ordered phase the situation is reversed, with $\mathcal{O}_{P,i,r} = 0$ and $\mathcal{O}_{xx,i,r} \neq 0$.

\section{The phase diagram}
\label{sec:phase_diagram}

\begin{figure}[t!]
    \centering
    \begin{minipage}{0.75\linewidth}
        \includegraphics[width=\linewidth]{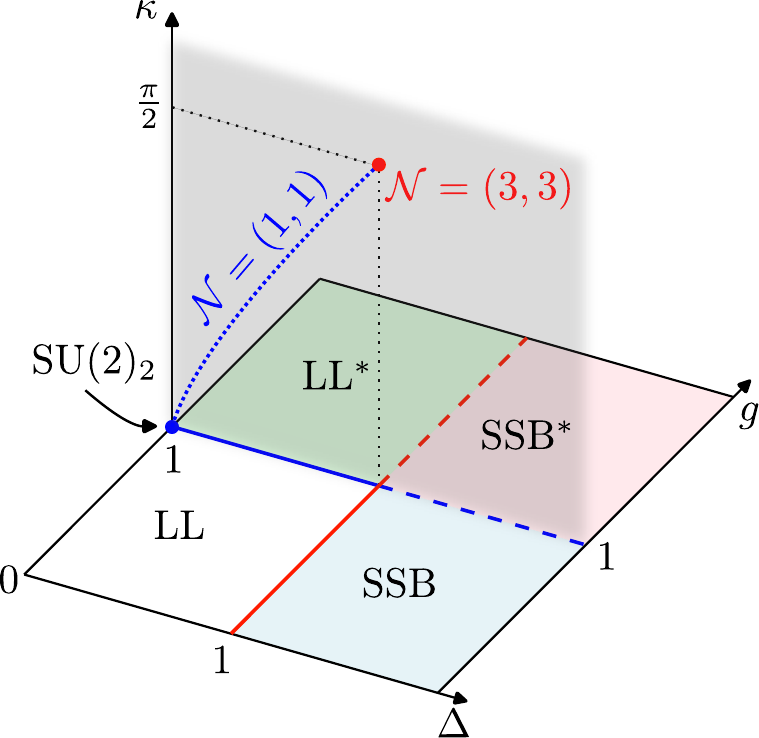}
    \end{minipage}
    \caption{\footnotesize (color online)
    Phase diagram of the Hamiltonian Eq.~\eqref{eq:H-in-Gauge-Invariant-Fields}.
    For any finite $\kappa>0$ and $|\Delta|<1$, conventional and gauged Luttinger liquid phases occur for $g<1$ (LL) and $g>1$ (LL$^{*}$), respectively.
    For $|\Delta|>1$, symmetry-broken phases appear: for $g<1$ (SSB) the $\mathbb{Z}_{2}^{C}$ symmetry is broken, while for $g>1$ (SSB$^{*}$) both $\mathbb{Z}_{2}^{C}$ and $\mathbb{Z}_{2}^{W}$ are spontaneously broken.
    Transitions along solid and dashed red lines are of BKT type, whereas those along solid and dashed blue lines are in the Ising universality class.
    The SSB–SSB$^{*}$ transition has central charge $c=\flatfrac{1}{2}$, and the LL–LL$^{*}$ line has $c=\flatfrac{3}{2}$ due to an additional gapless Luttinger liquid mode. Superconformal field theories (SUSY CFTs) appear on the dashed blue line, for $\kappa$ given by Eq.~\eqref{eq:SUSY-Condition}.}
    \label{fig:PhaseDiagramGaugeInvariant}
\end{figure}

In the previous section, we employed the transformation Eq.~\eqref{eq:nonlocal-mapping} to decompose the model into decoupled exactly-solvable XXZ and Ising sectors.
This allowed us to identify the locations and universality classes of the phase transition lines in the phase diagram in Fig.~\ref{fig:PhaseDiagramGaugeInvariant}.
Note, however, that the non-local character of the transformation obscures the interpretation of the phases of the original $\mathbb{Z}_{2}$ gauge theory, Eq.~\eqref{eq:Orthogonal-Metal}.
To clarify the nature of these phases, it is instructive to consider the following two limiting regimes of the model Eq.~\eqref{eq:H-in-Gauge-Invariant-Fields} expressed in the gauge-invariant formulation.

\subsection{\texorpdfstring{$g\to0$}{TEXT} regime}
In the $g \to 0$ limit, the $\Gamma$ term can be neglected and the spin sector fully polarizes along the $+z$ direction.
Consequently, we may set $Z_{j+\flatfrac{1}{2}} \to +1$ on all links.
Under this condition, the Hamiltonian in Eq.~\eqref{eq:H-in-Gauge-Invariant-Fields} reduces to an interacting spinless-fermion model:
\begin{equation}
\label{eq:g=0-limit}
    \begin{split}
        H_{0} = -& t \sum_{j}
        \left(
            c^{\dag}_{j}
            c^{\phantom{\dag}}_{j+1}
            + \text{h.c.}
        \right)
        \\
        +& V \sum_{j}
        \left( n_{j}-\frac{1}{2} \right)
        \left( n_{j+1}-\frac{1}{2} \right).
    \end{split}
\end{equation}
In this limit, two phases emerge: the LL phase with power-law correlations and continuously varying critical exponents, and a CDW phase with spontaneously broken $\mathbb{Z}_{2}^{C}$ symmetry.
They are separated by a BKT transition line, shown in Fig.~\ref{fig:PhaseDiagramGaugeInvariant} as the solid red line between the LL and SSB phases in the $\Delta$--$g$ plane for any finite $\kappa$.
\subsection{\texorpdfstring{$g\to\infty$}{TEXT} regime}
In the opposite limit $g\to\infty$, the $J$ term becomes negligible, and the following Gauss law is imposed energetically:
\begin{equation}
\label{eq:Emergent-Gauss-Law}
    \widetilde{G}_{j}=X_{j-\frac{1}{2}} (-1)^{n_{j}} X_{j+\frac{1}{2}}\to +1,
\end{equation}
which implies that the spin degrees of freedom effectively behave as an emergent $\mathbb{Z}_{2}$ gauge field.
In this regime, the Hamiltonian simplifies to \cite{Borla2020,kebrivc2023confinement}
\begin{equation}
\label{eq:g=infty-limit}
    \begin{split}
        H_{\infty} = -& t \sum_{j}
        \left(
            c^{\dag}_{j}
            Z_{j+\frac{1}{2}}
            c^{\phantom{\dag}}_{j+1}
            + \text{h.c.}
        \right)
        \\
        +& V \sum_{j}
        \left( n_{j} - \frac{1}{2} \right)
        \left( n_{j+1} - \frac{1}{2} \right).
    \end{split}
\end{equation}

As shown in~\cite{kebrivc2023confinement}, in this $\mathbb{Z}_2$ lattice gauge theory the system undergoes a transition from a gapless LL phase to a CDW-ordered phase via a BKT phase transition, marked by the red dashed line.
These phases are labeled LL$^{*}$ and SSB$^{*}$ in the $\Delta$–$g$ phase diagram on Fig.~\ref{fig:PhaseDiagramGaugeInvariant}.
A bosonic analogue of this model was recently investigated in Ref.~\cite{Su_Zeng_2024}.

The LL$^{*}$ phase differs from a conventional Luttinger liquid in that only states with even fermion parity are allowed.
This follows from multiplying the Gauss law in Eq.~\eqref{eq:Emergent-Gauss-Law} around the closed chain:
\begin{equation}
    P = \prod_{j = 1}^{L}
        \widetilde{G}_{j} = 1.
\end{equation}
Here, $P \equiv (-1)^{ \sum_{j} n_{j} }$ is the fermion parity operator, generating a $\mathbb{Z}_{2}^{P}$ subgroup of the global particle-number U(1) symmetry.
Although the parity-even constraint is strictly enforced only in the \( g \to \infty \) limit, it remains effectively valid throughout the entire LL\(^*\) phase (\( g > 1 \)) in the low-energy sector.
The same applies to the transition line separating LL$^{*}$ and SSB$^{*}$, which we denote by BKT$^{*}$.
In the CFT framework, the only distinction from the standard BKT transition is that all operators that do not commute with fermion parity must be removed from the operator content.

\begin{figure}[t!]
    \centering
    \begin{minipage}{\linewidth}
        \includegraphics[width=\linewidth]{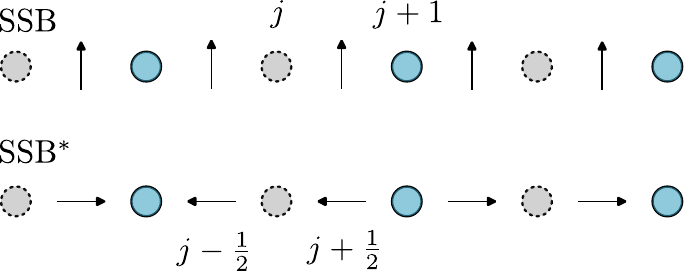}
    \end{minipage}
    \caption{\footnotesize (color online) Schematic illustration of the ground states in the $\mathbb{Z}_{2}^{C}$-breaking SSB phase ($\Delta > 1$, $g < 1$) and $\mathbb{Z}_{2}^{C}\times\mathbb{Z}_{2}^{W}$-breaking SSB$^{*}$ ($\Delta > 1$, $g > 1$) phase.
    }
    \label{fig:SSB}
\end{figure}

In the SSB$^{*}$ phase, once the $\mathbb{Z}_{2}^{C}$-breaking CDW pattern is formed, the emergent Gauss law forces the $x$-components of spins to point either toward or away from the fermion-occupied sites,  thereby also spontaneously breaking the $\mathbb{Z}_{2}^{W}$ symmetry and leading to a four-fold degenerate ground state. 
Fig.~\ref{fig:SSB} shows a schematic representation of the ground-state configurations in the SSB and SSB$^{*}$ phases.

\section{Numerical results}
\label{sec:numerics}
To shed more light on the phase diagram, we performed DMRG simulations of the orthogonal fermion model Eq.~\eqref{eq:H-in-Gauge-Invariant-Fields}. 
Using TeNPy libraries~\cite{tenpy}, we set $t=1$ and explore a wide range of the parameter space $(\Delta,g,\kappa)$ of the Hamiltonian in Eq.~\eqref{eq:H-in-Gauge-Invariant-Fields}, determine the ground state, and compute all relevant correlation functions, Eqs.~\eqref{eq:Structure_factor} and~\eqref{eq:O_P,O_xx}.
We also evaluate the bipartite entanglement entropy of the ground state and obtain the central charge using the Calabrese-Cardy formula~\cite{Calabrese2004}.

To implement the Hamiltonian Eq.~\eqref{eq:H-in-Gauge-Invariant-Fields} within DMRG, we consider a finite chain of doubled length, where complex fermions occupy the even sites and Pauli spin-$\flatfrac{1}{2}$ operators occupy the odd sites.
In this construction, the numbers of fermionic and spin sites are both fixed to $L$, with $L$ chosen to be an even integer.
We further assume that the left boundary of the doubled chain begins with a fermionic site, while the right boundary ends with a spin site.
For sufficiently large systems, one can either add an extra spin site at the left edge or remove the rightmost spin site without influencing the bulk physics.

Since multi-critical points are more subtle and susceptible to finite-size effects, we employ both DMRG and iDMRG approaches to study these criticalities.
In particular, we use iDMRG~\cite{itensor} to extract central charges at the ${\mathcal{N}=(3,3)}$ point and at the SU(2)$_{2}$ point, while DMRG~\cite{tenpy} is used to determine fermionic and bosonic velocities.

Given that the lattice model factorizes into XXZ and QI parts, we focus on the order parameters in the XXZ and QI sectors in Secs.~\ref{subsec:XXZ} and~\ref{subsec:QI}, respectively.
The results for the central charges of the full model are discussed in Sec.~\ref{subsec:central_charges}.
The SUSY points will be analyzed in detail in Sec.~\ref{sec:SUSY}.
\subsection{XXZ sector}
\label{subsec:XXZ}
\begin{figure}[ht!]
    \centering
    \begin{minipage}{\linewidth}
        \includegraphics[width=0.75\linewidth]{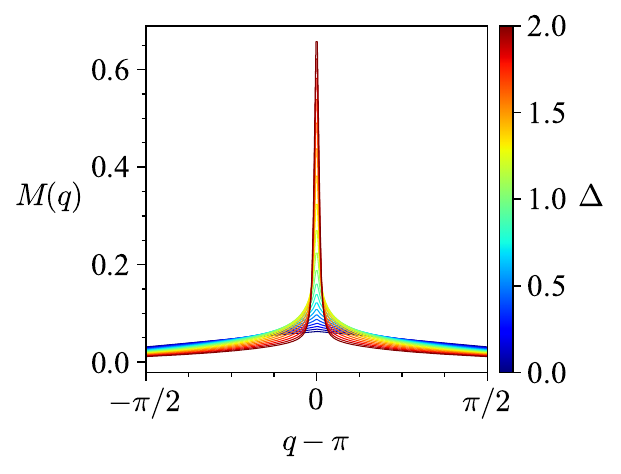}
    \end{minipage}
    \caption{\footnotesize (color online) Numerically obtained $M(q)$ for $L=256$, for several values of $0 \leq \Delta \leq 2$ at fixed $\kappa = 1$.
    Here we set $g = 0.5$, although the results are identical for any $g$, as $M(q)$ is insensitive to the properties of the QI sector of the model.
    For small anisotropies $0< \Delta < 1$, $M(q)$ increases linearly with $q$. In contrast, for larger anisotropies $1 < \Delta$ we observe a pronounced peak at $q = \pi = 2k_{F}$, signaling the emergence of CDW order in the XXZ sector.}
    \label{fig:tructure_Function_vs_Delta}
\end{figure}

In this subsection, we focus on the correlators associated with the XXZ physics.
Specifically, we compute the structure factor $M(q)$,  
\begin{equation}
    \label{eq:Structure_factor-mapped}
    M(q) = \frac{1}{L}
        \sum_{j,j'} e^{iq(j-j')}
        \left\langle
            \left\langle n_{j} n_{j'} \right\rangle
        \right\rangle
        = \frac{S(q)}{4},
\end{equation}
obtained from Eq.~\eqref{eq:Structure_factor} using the non-local map in Eqs.~\eqref{eq:nonlocal-mapping}, and verify that the numerical results are consistent with the theoretical predictions.
In Fig.~\ref{fig:tructure_Function_vs_Delta}, we show the profile of $M(q)$ for various values of $\Delta$.

\begin{figure}[ht!]
    \centering
    \begin{minipage}{\linewidth}
        \includegraphics[width=0.8\linewidth]{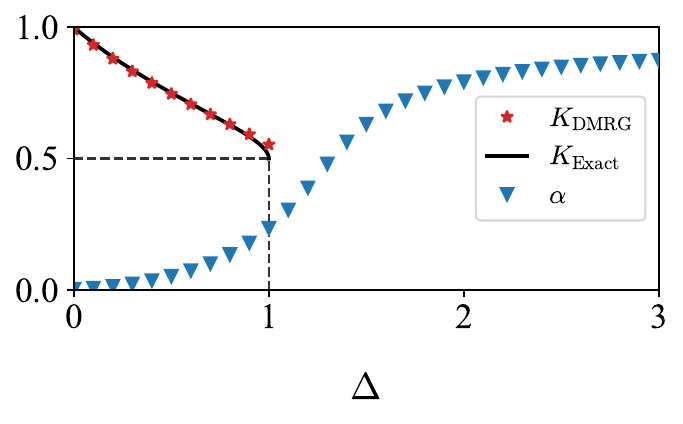}
    \end{minipage}
    \caption{\footnotesize (color online) Numerically obtained values of $K$ and $\alpha$ over a wide range of $\Delta$ at fixed system size $L=256$.
    The Luttinger parameter $K$ is extracted by fitting the structure factor at momenta $q/\pi < 0.1$ to the form $f(q) = Kq/2\pi$, yielding $K_{\text{DMRG}}$ in excellent agreement with the exact result. 
    The exponent $\alpha$ is obtained from a fit of $M(\pi)$ to $f(L) = AL^{\alpha}$ for $L\in[32,64,128,256]$.
    Over this interval of system sizes, $\alpha$ gradually converges towards the expected value $\alpha=1$, signaling a tendency towards true long-range CDW order in the corresponding parameter regime.}
    \label{fig:K_and_alpha_vs_Delta}
\end{figure}

For small $\Delta$, $M(q)$ displays a linear dependence on $q$, in agreement with the Luttinger liquid theory.
Using the scaling form in Eq.~\eqref{eq:Structure_factor_in_LL}, we numerically determine the Luttinger parameter of the model, as shown on Fig.~\ref{fig:K_and_alpha_vs_Delta}.
For $\Delta < 1$ the extracted values of $K(\Delta)$ are in good qualitative agreement with Eq.~\eqref{eq:LL-Parameter}.
On the other hand, as one approaches $\Delta \to 1^{-}$, we observe a small deviation from the exact value of $K=\flatfrac{1}{2}$, due to finite size effects originating from a marginally irrelevant cosine-perturbation in Eq.~\eqref{eq:XXZ-Action}.

For larger values of $\Delta>1$, the density-density interaction locks the system into a CDW phase with characteristic momentum $q = 2k_{F} = \pi$.
In this regime, a pronounced peak in $M(q)$ is expected for momenta near $2k_{F}$, which is clearly visible on Fig.~\ref{fig:tructure_Function_vs_Delta}.
For a true long-range CDW order, one expects a scaling behavior $M(2k_{F}) \sim L^{\alpha}$ with an exponent $\alpha \approx 1$.
As shown on Fig.~\ref{fig:K_and_alpha_vs_Delta}, within the accessible system sizes, the extracted $\alpha$ gradually converges toward the expected value $\alpha = 1$ as $\Delta$ increases.
\subsection{QI sector}
\label{subsec:QI}
\begin{figure}[t!]
    \centering
    \begin{minipage}{\linewidth}
        \includegraphics[width=\linewidth]{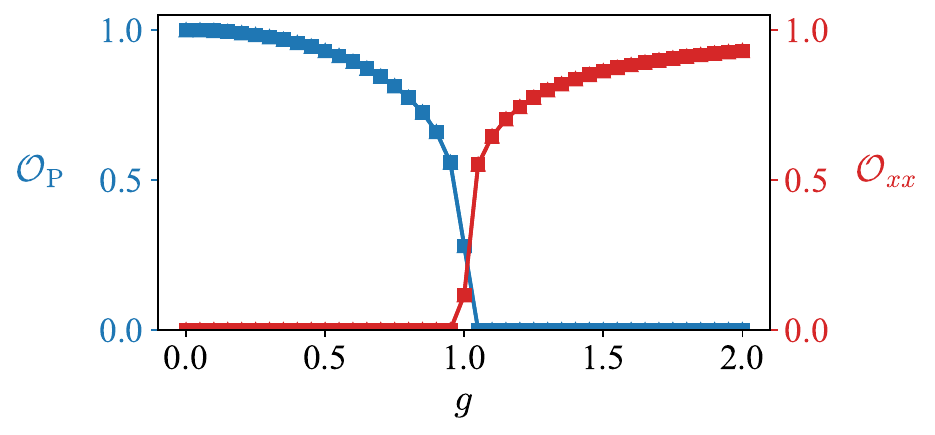}
    \end{minipage}
    \caption{\footnotesize (color online) Numerical results for the expectation values of the order and disorder parameters. We fix the system size to $L=256$ and $\Delta=0.5$.
    Since the definitions in Eqs.~\eqref{eq:O_P,O_xx} depend on bulk indices $i$ and $r$, we choose $i=L/4$ and $r=L/2$ to minimize finite-size effects, i.e., we evaluate $\mathcal{O}_{P}$ and $\mathcal{O}_{xx}$ in the central part of the chain.
    As $g$ varies, these order and disorder parameters clearly signal the expected Ising transition at $g=1$.}
    \label{fig:QI_topology_vs_g}
\end{figure}

In this subsection, we report numerical results for the correlators relevant to the QI sector.
As argued in Sec. \ref{sec:analytics}, for any $\Delta$, the observables $\mathcal{O}_{P,r}$ and $\mathcal{O}_{xx,r}$, which in terms of the gauge invariant degrees of freedom admit the string-order forms:
\begin{equation}
    \label{eq:O_P,O_xx-mapped}
    \begin{split}
        \mathcal{O}_{P,i,r}
            &=
            \left\langle
                \prod^{i+r}_{j=i} Z_{j+\frac{1}{2}}
            \right\rangle, 
        \\
        \mathcal{O}_{xx,i,r}
            &=
            \left\langle
                X_{i+\frac{1}{2}}
                \left[ \prod^{i+r-1}_{j=i+1}(-1)^{n_{j}} \right]
                X_{i+r+\frac{1}{2}}
            \right\rangle,
    \end{split}
\end{equation}
should indicate the phase transition across the critical line at $g=1$.
From the behavior of $\mathcal{O}_{P,r}$ and $\mathcal{O}_{xx,r}$, we can determine whether the system is in the ordered or disordered phase of the dual Ising model.
The numerical results for $L=256$ are shown on Fig.~\ref{fig:QI_topology_vs_g}.
We present data for $\Delta = 0.5$ only, since the case $\Delta = 1.5$ is identical.
These order parameters correctly identify the region $g<1$ as disordered and $g>1$ as ordered phases, and they also accurately locate the critical point at $g \approx 1$.

\begin{figure}[h!]
    \centering
    \begin{minipage}{\linewidth}
        \includegraphics[width=0.75\linewidth]{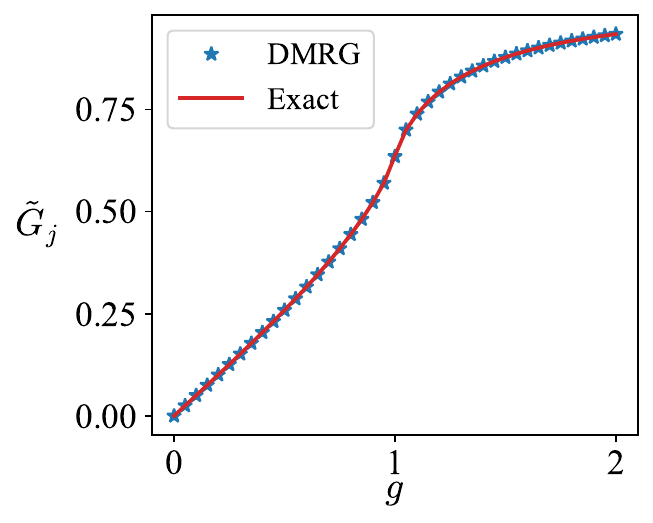}
    \end{minipage}
    \caption{\footnotesize (color online) Comparison of the numerical results for $L=256$ and exact result Eq.~\eqref{eq:Soft_vs_Strong_Gauge_2} for infinite system. To reduce the finite size effects, we plot $\left\langle X_{j+\flatfrac{1}{2}}(-1)^{n_{j}}X_{j+1+\flatfrac{1}{2}} \right\rangle$ with $j=L/2$.}
    \label{fig:Gauge_vs_g}
\end{figure}

Additionally, from Eq.~\eqref{eq:H_QI} one can evaluate the nearest-neighbor $x$–$x$ correlator:
\begin{align}
\nonumber
        \langle \tau^{x}_{j} \rangle
            &= \left\langle
                \mu^{x}_{j-\frac{1}{2}}\mu^{x}_{j+\frac{1}{2}} 
            \right\rangle 
            \\
\label{eq:Soft_vs_Strong_Gauge_1}
            &= \left\langle
                X_{j-\frac{1}{2}}(-1)^{n_{j}}X_{j+\frac{1}{2}} 
            \right\rangle
            = \tilde{G}_{j}\left(g\right),
\end{align}
which provides a natural diagnostic of the emergent gauge symmetry associated with the Gauss-law constraint in Eq.~\eqref{eq:Emergent-Gauss-Law}.
In the thermodynamic limit, we obtain \cite{Santoro2024}
\begin{equation}
    \label{eq:Soft_vs_Strong_Gauge_2}
    \tilde{G}_{j}(g) = \frac{1}{\pi} \int^{\pi}_{0} dk \frac{g- \cos k}{\sqrt{1+g^2-2 g \cos k}} .
\end{equation}
For $g \gg 1$, this yields $\tilde{G}_{j} \approx 1 - 1/(2g^2)$, indicating the emergent Gauss constraint.
In contrast, for $0< g \ll 1$, we find $\tilde{G}_{j} \approx g/2$. In Fig.~\ref{fig:Gauge_vs_g} we compare the numerically obtained values of $\left\langle X_{j+\flatfrac{1}{2}}(-1)^{n_{j}}X_{j+1+\flatfrac{1}{2}} \right\rangle$ with the exact expression, Eq.~\eqref{eq:Soft_vs_Strong_Gauge_2}.

\subsection{Central charges}
\label{subsec:central_charges}
\begin{figure}[t!]
    \centering
    \begin{minipage}{\linewidth}
        \includegraphics[width=\linewidth]{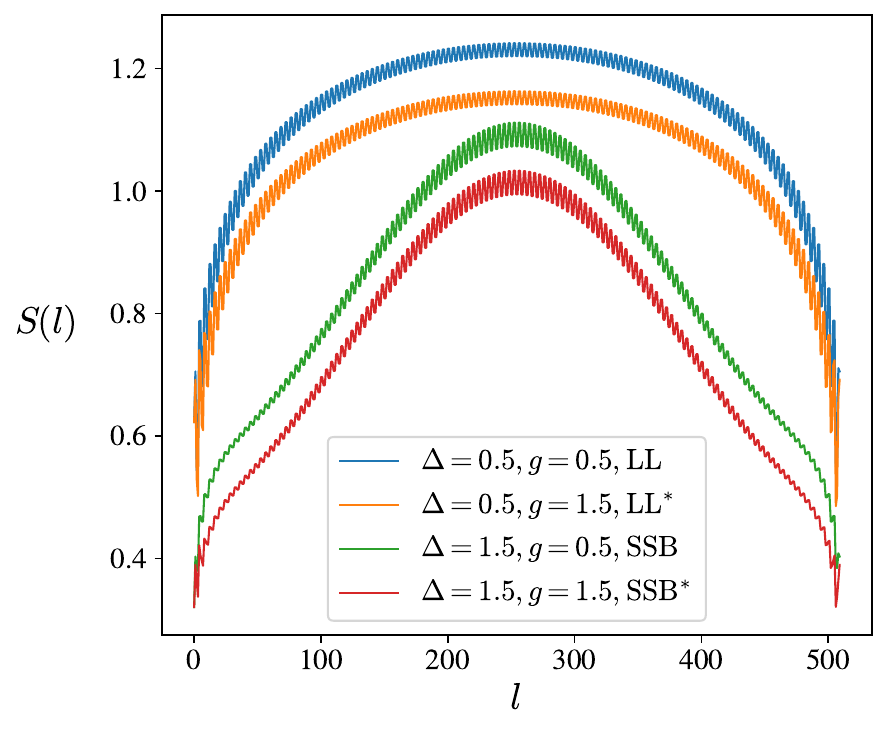}
    \end{minipage}
    \caption{\footnotesize (color online) Bipartite entanglement entropy $S(l)$ as a function of the bipartition size $l$, with $L=256$. We fix $\kappa = 1$ and show 4 different parameter regimes. For LL and LL$^{*}$ we observe logarithmic scaling of $S(l)$, typical for a CFT. The extracted central charges are $c \approx 0.96$ and $c \approx 0.93$ for LL and LL$^{*}$, respectively. For SSB and SSB$^{*}$, the bell-shaped behavior of $S(l)$ indicates the absence of underlying CFT description. The additional oscillations are $S(l)$ is due to residual entanglement between fermions and spin fields.}
    \label{fig:central_charge}
\end{figure}

To extract the central charge, we compute the bipartite von Neumann entanglement entropy of a block of $l$ sites,
\begin{equation}
    S(l) = -\mathrm{Tr}\big[\rho_l \ln(\rho_l)\big],
\end{equation}
where $\rho_l$ is the reduced density matrix of the $l$-site subsystem.
The central charge $c$ of the underlying CFT governing a critical phase is obtained from the scaling of $S(l)$ with subsystem size $l$.
For a 1D critical ground state described by a CFT, the finite-size scaling for a bipartition of size $l$ in a chain of length $L$ with OBC is
\begin{equation}
    \label{eq:calabrese}
    S(l, L) = \frac{c}{6} \ln\!\left[\frac{L}{\pi} \sin\!\left(\frac{\pi l}{L}\right)\right] + c',
\end{equation}
where $c'$ is a nonuniversal constant and $c$ is the central charge~\cite{Callan1994, Vidal2003, Calabrese2004}.
We compute the von Neumann entropy for a subsystem of length $l$ that contains both fermionic and spin sites.
Fitting the scaling of $S(l)$ \textit{vs.} $l$ then, in the language of the previous section, yields the total central charge
\begin{equation}
    c = c_{\text{B}} + c_{\text{F}},
\end{equation}
with $c_{\text{B}} = 1$ ($c_{\text{B}} = 0$) if the XXZ sector is critical (gapped) and $c_{\text{F}} = 1/2$ ($c_{\text{F}} = 0$) if the QI sector is critical (gapped).
The entanglement scaling for a system of $L=256$ fermionic and $256$ spin sites is shown on Fig.~\ref{fig:central_charge}.
For LL and LL$^{*}$ we observe the expected logarithmic growth of $S(l)$, with central charges $c \approx 0.96$ and $c \approx 0.93$, respectively.
In contrast, the absence of logarithmic scaling in the SSB and SSB$^{*}$ phases indicates gapped phases.

\subsection{The SUSY points}
\label{sec:SUSY}
\begin{figure}[t!]
    \centering
    \begin{minipage}{\linewidth}
        \includegraphics[width=\linewidth]{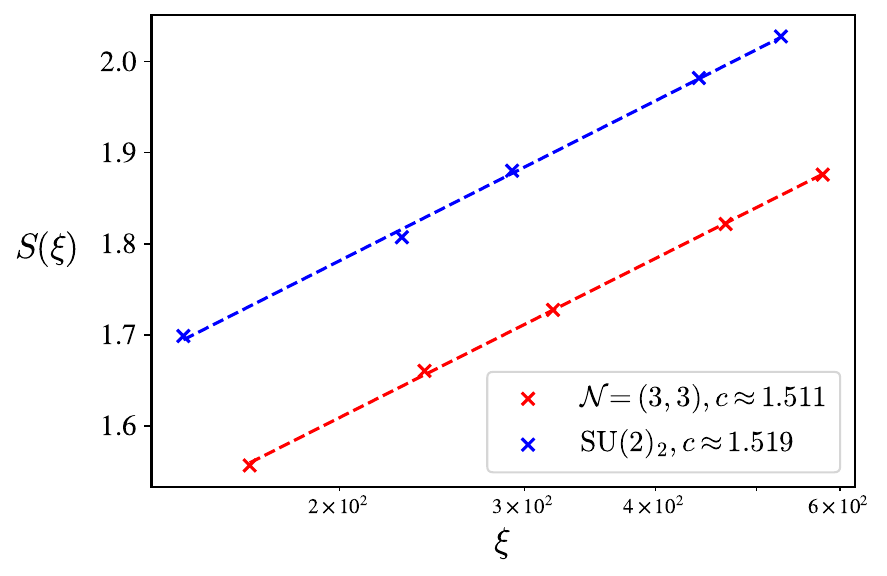}
    \end{minipage}
    \caption{\footnotesize (color online) Entanglement entropy $S$ vs the correlation length $\xi$ from  iDMRG simulations at SU(2)$_{2}$ and $\mathcal{N}=(3,3)$.
    The values of the corresponding central charges are in very good agreement with the exact value of $c=\flatfrac{3}{2}$.}
    \label{fig:SUSY_points_central_charge}
\end{figure}

To reliably determine the central charge of the lattice model of Eq.~\eqref{eq:H-in-Gauge-Invariant-Fields} at the SUSY points, we employ a combination of finite DMRG and iDMRG methods using TenPy and ITensor numerical tools ~\cite{tenpy, itensor}. 
In particular, for the SUSY CFT points of the phase diagram, represented by the SU(2)$_2$ and $\mathcal{N}=(3,3)$ theories, we rely on iDMRG.
There, we obtain the ground state for a sequence of maximum bond dimensions $\chi$ and, for each $\chi$, compute the associated correlation length $\xi_{\chi}$ and the entanglement entropy $S_{\chi}$.
The conformal invariance then imposes the scaling relation \cite{Calabrese2004,PhysRevB.78.024410, PhysRevLett.102.255701}
\begin{equation}
    \label{eq:S_xi}
    S_{\chi} \approx \frac{c}{6}
        \log\!\left( \frac{\xi_{\chi}}{a_{0}} \right) + \text{const},
\end{equation}
with $a_{0}=1$ the lattice spacing.
From this scaling, we extract the central charges at the SU(2)$_2$ and $\mathcal{N}=(3,3)$ SUSY points; see Fig.~\ref{fig:SUSY_points_central_charge}.
In both cases, the resulting $c$ is very close to the expected value $\flatfrac{3}{2}$, consistent with the critical XXZ and QI sectors.

The simultaneous criticality of the XXZ and QI theories is not, by itself, sufficient to establish supersymmetry.
To identify the critical point as supersymmetric, one must additionally require the excitation velocities of the bosonic (XXZ) and fermionic (QI) sectors to be  equal.

Since after the non-local mapping from Sec. \ref{sec:analytics} the full model factorizes into XXZ and QI contributions, the ground-state energy of a finite system can be written as
\begin{equation}
    E_{0}(L,N,W) = E_{\text{XXZ}}(L,N) + E_{\text{QI}}(L,W),
\end{equation}
where $E_{\text{XXZ}}(L,N)$ is the ground-state energy of an open XXZ chain with $L$ sites and $N$ fermions, and $E_{\text{QI}}(L,W)$ is the ground-state energy of an open QI chain with $L$ sites in the parity sector $W$, given in Eq.~\eqref{eq:W-Transformation}.
To obtain the Ising velocity $u$, we consider the Ising energy gap between the ground states of opposite parity:
\begin{equation}
    \label{eq:Delta_QI}
    \Delta_{\text{QI}}(L) = E_{0}(L,N,+1) - E_{0}(L,N,-1),
\end{equation}
which at criticality scales as \cite{Ginsparg1988}
\begin{equation}
    \Delta_{\text{QI}}(L) = \frac{\pi u}{2L}.
\end{equation}

To determine the XXZ velocity, Eq.~\eqref{eq:LL-Velocity}, we evaluate the charge gap
\begin{equation}
    \label{eq:Delta_XXZ}
    \begin{split}
        \Delta_{\text{XXZ}}(L) = E_{0}(L,N+1,W) + &E_{0}(L,N-1,W) 
        \\
       -2&E_{0}(L,N,W),
    \end{split}
\end{equation}
which in the leading order follows the scaling law:
\begin{equation}
    \Delta_{\text{XXZ}}(L)
        = \frac{1}{ L \chi_{\text{c}} }
        = \frac{\pi v}{L K},
\end{equation}
where $\chi_{\text{c}}$ is the charge compressibility~\cite{Giamarchi2003}, and the Luttinger parameter $K$ is defined in Eq.~\eqref{eq:LL-Parameter}.

For a simple representative of the $\mathcal{N} = (1,1)$ SUSY point, we take SU(2)$_{2}$ criticality.
At this point, the velocities should be
\begin{equation}
    v = u = 2t,
\end{equation}
while for the $\mathcal{N}=(3,3)$ SUSY point
\begin{equation}
    v = u = \pi t,
\end{equation}
as expected from Eq.~\eqref{eq:LL-Velocity}.
\begin{figure}[t!]
    \centering
    \begin{minipage}{\linewidth}        \includegraphics[width=\linewidth]{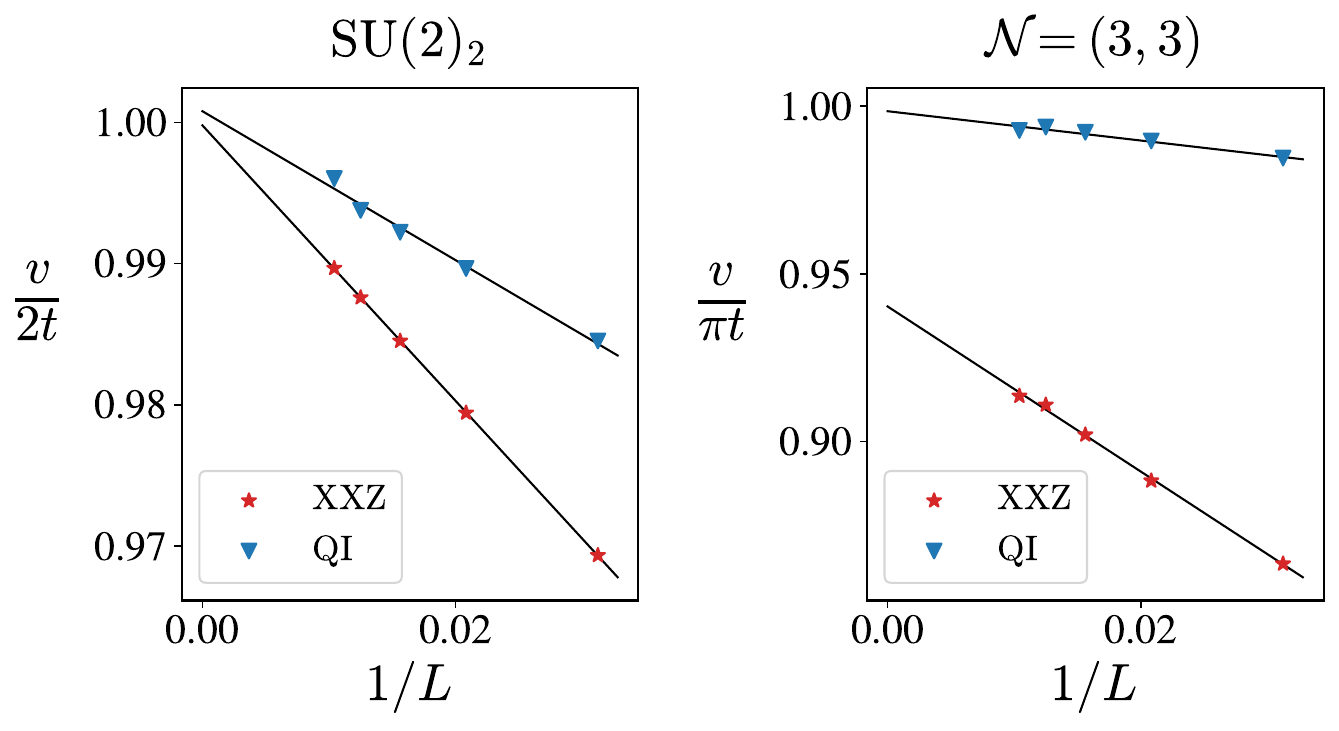}
    \end{minipage}
    \caption{\footnotesize (color online) By sampling QI and XXZ gaps for finite system sizes and using Eq.~\eqref{eq:Delta_QI} and Eq.~\eqref{eq:Delta_XXZ}, we numerically extract the fermionic and bosonic velocities in the model Eq.~\eqref{eq:H-in-Gauge-Invariant-Fields} at two SUSY points.
    By performing the fitting procedure explained in the main text, we extrapolate our results to infinite system size.
    In both cases, we see a clear trend of the velocities approaching the theoretically predicted values of $2t$ for SU(2)$_{2}$ and $\pi t$ for $\mathcal{N}=(3,3)$.
    At the $\mathcal{N}=(3,3)$ point, however, the XXZ sector lies at a BKT transition, where finite-size effects are particularly strong; as a result, our numerical estimate shows a slight deviation from the expected value.}
    \label{fig:susy_velocities}
\end{figure}

Since the velocities $u$ and $v$ can be inferred from the Ising and charge excitation gaps, which depend only on the total energy, they are accessible numerically through a finite DMRG.
Fig.~\ref{fig:susy_velocities} shows DMRG results for the model Eq.~\eqref{eq:H-in-Gauge-Invariant-Fields} for the system sizes $L=\{32,48,64,96\}$ at the SU(2)$_2$ and $\mathcal{N}=(3,3)$ points.
We extract the thermodynamic-limit velocity $v_{\infty}$ by fitting the data to $v(L)=v_{\infty}+a/L$ and plotting $v(L)$ versus $1/L$.
At the SU(2)$_2$ point, both velocities converge to the expected value $2t$ and agree with each other with high precision.
At the $\mathcal{N}=(3,3)$ point, the QI velocity tends to $\pi t$, while the extrapolated XXZ velocity deviates from the expected value by about $6\%$.
This small discrepancy between numerics and the exact result, Eq.~\eqref{eq:LL-Velocity}, at $\Delta=1$ arises because at the $\mathcal{N}=(3,3)$ point the XXZ chain sits exactly at the BKT transition, where a marginally irrelevant operator induces strong finite-size corrections.
\section{Conclusion and discussions}
\label{sec:conclusion}
In this work, we analyzed the phase diagram of a one-dimensional orthogonal metal model.
The system consists of a 1D chain of spinless fermions and spin-$\flatfrac{1}{2}$ matter fields, minimally coupled to a $\mathbb{Z}_{2}$ gauge field with a hard Ising gauge constraint.

The model is exactly solvable: first, employing a standard resolution of the Gauss law, we recast the original lattice Hamiltonian in terms of gauge-invariant fermionic and spin-$\flatfrac{1}{2}$ variables.
A non-local and non-linear transformation, Eq.~\eqref{eq:nonlocal-mapping}, then maps the system onto the decoupled spin-$\flatfrac{1}{2}$ XXZ Heisenberg chain (with $z$-anisotropy $\Delta$) and a transverse-field quantum Ising (QI) chain (with dimensionless exchange $g$).
Since both models are exactly solvable, we can precisely describe the structure of the phase diagram of the original Hamiltonian across the parameter space.

In the three-dimensional parameter space $(\Delta, g, \kappa)$, we identified several gapless regimes, alongside gapped phases that exhibit spontaneous symmetry breaking. 
For weak fermionic repulsion ($|\Delta| < 1$) and with strong transverse $Z$--field in the spin sector ($g < 1$), the fermions form an ordinary gapless Luttinger liquid (LL).
In the weak transverse $Z$--field regime ($g > 1$), the fermions remain gapless, but the Gauss law Eq.~\eqref{eq:Emergent-Gauss-Law} emerges at low energies. This effectively restricts the system to the even-fermion-parity sector, giving rise to the LL$^{*}$ phase. 
For strong repulsion ($\Delta > 1$), the fermions develop a charge-density-wave (CDW) order with a dynamically generated gap, spontaneously breaking the particle-hole $\mathbb{Z}_{2}^{C}$ symmetry. In the $g < 1$ regime, this results in a doubly degenerate ground state (SSB). 
In the $g > 1$ regime, the CDW order intertwines with the emergent gauge constraint, forcing the spins to simultaneously break the magnetic $\mathbb{Z}_{2}^{W}$ symmetry, which leads to a four-fold degenerate ground state (SSB$^{*}$). 
At $\Delta = 1$, the fermionic sector exhibits a BKT transition, while at $g = 1$ the spin sector undergoes a quantum phase transition belonging to the Ising universality class.

In addition to these criticalities, the specific values of $\kappa$ are characterized by additional emerging symmetries.
Namely, when the excitation velocities of a critical Ising model and the XXZ theory become equal, one encounters supersymmetric conformal field theories of various universality classes.
For the velocities of bosonic (XXZ) and fermionic (QI) to coincide, one has to fine tune the parameters to $g=1$ and ${\kappa = \kappa_{\text{SUSY}} =\pi\sqrt{1-\Delta^{2}}/2\arccos{\Delta}}$.
Along the SUSY line, $\kappa = \kappa_{\text{SUSY}}$, for any $0 \leq \Delta < 1$ the theory is described by a $\mathcal{N} = (1,1)$ superconformal field theory.
Exactly at $\Delta = 0$, the criticality belongs to SU(2)$_{2}$ WZNW universality class.
On the other hand, at $\Delta = 1$ the effective low-energy theory of the model can be unified into a $\mathcal{N} = (3,3)$ SUSY CFT.

To validate the theoretical predictions, we investigated the microscopic Hamiltonian, Eq.~\eqref{eq:H-in-Gauge-Invariant-Fields}, using finite and infinite DMRG.
For system sizes up to $L=256$, the numerical data are in excellent agreement with the analytical results.
In the LL and LL$^{*}$ phases, the Luttinger parameter $K$ obtained from DMRG matches the exact Bethe ansatz values, and the onset of charge-density-wave order in the SSB and SSB$^{*}$ phases is accurately reproduced.
Correlation functions relevant to the QI sector are likewise determined with high precision.
From the scaling of the von Neumann entanglement entropy at various points in the phase diagram, we extract central charges consistent with the underlying CFTs.
Crucially, the factorization of the model into independent XXZ and QI Hamiltonians allows for the direct determination of bosonic and fermionic velocities.
The finite-size DMRG at the SU(2)$_{2}$ and ${\mathcal{N} = (3,3)}$ SUSY points 
shows that these velocities are in good agreement and converge to the predicted values ${v=2t}$ (free-fermion limit) at SU(2)$_{2}$ and ${v=\pi t}$ (Bethe ansatz) at ${\mathcal{N} = (3,3)}$.

Our exact results provide a natural starting point to explore the stability of various phases against additional perturbations. Namely, a particularly interesting direction is to investigate the deformations of the original model at the SUSY points, which could reveal whether the $\mathcal{N} = (1,1)$ and $\mathcal{N} = (3,3)$ regimes can host extended supersymmetric phases or if they are unstable isolated criticalities. Furthermore, while the current work relies on deconfined gauge fields, introducing gauge fields dynamics into the Hamiltonian of the orthogonal metal -- specifically, electric field fluctuations -- would be an interesting extension of our model. This perturbation couples the QI and XXZ sectors to each other in a very non-local fashion and thus can lead to new exotic phases. Additionally, such electric fields mediate confinement of the fractionalized degrees of freedom and it would be of interest to investigate quantum dynamics in this model including the string-breaking phenomenon.

\acknowledgements

The authors thank T. Chanda, M. Dalmonte, G. Japaridze, E. Jmukhadze, and A. Nersesyan for fruitful discussions.
We also thank E. Antadze for useful suggestions during the implementation of the DMRG code.
B.B. acknowledges funding from the Wenner-Gren Foundations (project number UPD2024-0111).
M.T. acknowledges funding by the Deutsche Forschungsgemeinschaft (DFG, German Research Foundation) under Projektnummer 277101999 – TRR 183, and under Germany’s Excellence Strategy – Cluster of Excellence Matter and Light for Quantum Computing (ML4Q) EXC 2004/2 – 390534769.
S.M. is supported by Vetenskapsr{\aa}det (grant number 2021-03685) and Nordita.
\appendix

\section{Mapping to decoupled XXZ and QI models}
\label{app:XXZ-QI_mapping}
In this Appendix, we provide the derivation of the mapping between the original orthogonal metal model, Eq.~\eqref{eq:H-in-Gauge-Invariant-Fields}, and the decoupled Hamiltonian in Eq.~\eqref{eq:Decoupled-Model}.
First, we apply the following Jordan–Wigner (JW) transformation to the fermionic sector of Eq.~\eqref{eq:H-in-Gauge-Invariant-Fields},
\begin{equation}
    \begin{split}
        c_{j} = \frac{s^{x}_{j}-\mathrm{i}s^{y}_{j}}{2}
            \prod_{l<j} (-s^{z}_{l}),
        \quad
        n_{j} = \frac{1}{2} s^{z}_{j} + \frac{1}{2},
    \end{split}
\end{equation}
and rename the spins on links,
\begin{equation}
    s^{x}_{j-\frac{1}{2}} \equiv X_{j-\frac{1}{2}},
    \quad
    s^{z}_{j-\frac{1}{2}} \equiv Z_{j-\frac{1}{2}},
\end{equation}
where Pauli matrices $s^{\alpha}_{k}$, $\alpha=x,y,z$ live in the extended lattice of sites and links $k=\flatfrac{1}{2},1,\flatfrac{3}{2},\cdots,L$. In terms of these degrees of freedom, we obtain the following spin model:
\begin{equation}
    \begin{split}
        H =-& \frac{t}{2} \sum_{j}
        \left(
            s^{x}_{j}
            s^{z}_{j+\frac{1}{2}}
            s^{x}_{j+1}
            +
            s^{y}_{j}
            s^{z}_{j+\frac{1}{2}}
            s^{y}_{j+1}
        \right)
        \\
        + & \frac{V}{4} \sum_{j}
        s^{z}_{j} s^{z}_{j+1}
        \\
        -&\Gamma \sum_{j}
        s^{x}_{j-\frac{1}{2}} s^{z}_{j} s^{x}_{j+\frac{1}{2}}
        -J \sum_{j} s^{z}_{j+\frac{1}{2}},
    \end{split}
\end{equation}
featuring next-to-nearest-neighbor three-spin couplings with amplitudes $t$ and $\Gamma$.
Applying now the JW transformation on the extended lattice,
\begin{equation}
    \begin{split}
        s^{x}_{k} &=
            \left( a^{\dag}_{k} + a^{\phantom{\dag}}_{k} \right)
            \exp\left(
                \mathrm{i} \pi
                \sum_{l<k} a^{\dag}_{l} a^{\phantom{\dag}}_{l}
            \right),
        \\
        s^{y}_{k} &=
            -\mathrm{i}
            \left( a^{\dag}_{k}-a^{\phantom{\dag}}_{k} \right)
            \exp\left(
                \mathrm{i} \pi
                \sum_{l<k} a^{\dag}_{l} a^{\phantom{\dag}}_{l}
            \right),
        \\
        s^{z}_{k} &= 2 n_{k}-1,
        \quad
        k=\frac{1}{2},1,\frac{3}{2},\cdots,L,
    \end{split}
\end{equation}
one sees explicitly that the spinless fermions living on the sites and on the links mutually decouple,
\begin{align}
    &\begin{aligned}
    \label{eq:XXZ-in-spinless-fermions}
        H_{\text{XXZ}} = ~ &t \sum_{j}
            \left(
                a^{\dag}_{j}
                a^{\phantom{\dag}}_{j+1}
                + \text{h.c.}
            \right)
            \\
            +& V \sum_{j}
            \left( n_{j}-\frac{1}{2} \right)
            \left( n_{j+1}-\frac{1}{2} \right),
    \end{aligned}
    \\
    &\begin{aligned}
    \label{eq:QI-spinless-fermions}
        H_{\text{QI}} = ~ & \Gamma \sum_{j}
            \left(
                a^{\dag}_{j-\frac{1}{2}} -
                a^{\phantom{\dag}}_{j-\frac{1}{2}}
            \right)
            \left(
                a^{\dag}_{j+\frac{1}{2}} +
                a^{\phantom{\dag}}_{j+\frac{1}{2}}
            \right)
            \\
            -&J \sum_{j} (2 n_{j+\frac{1}{2}}-1).
    \end{aligned}
\end{align}
One clearly recognizes spinless fermion realizations of the spin-$\flatfrac{1}{2}$ Heisenberg XXZ chain in Eq.~\eqref{eq:H_XXZ} and of the quantum Ising model in Eq.~\eqref{eq:H_QI}, respectively. This can be mathematically obtained via the final JW transformations on the fermions living on the sites and on the links independently:
\begin{equation}
    \begin{split}
        a_{m} &= \frac{\mu^{x}_{m}-\mathrm{i}\mu^{y}_{m}}{2}
            \prod_{l<m} (-\mu^{z}_{l}),
        \quad
        n_{m} = \frac{1}{2} + \frac{1}{2} \mu^{z}_{m},
    \end{split}
\end{equation}
where $m$ and $l$ run over sites ($m,l=1,2,\dots,L$) or over links ($m,l=\flatfrac{1}{2},\flatfrac{3}{2},\dots,L-\flatfrac{1}{2}$).
Combining all of the above into a single composite transformation gives us the total transformation Eq.~\eqref{eq:nonlocal-mapping}.

\section{Review of CFTs}
\label{appendix:CFTs}
In the $1+1$-dimensional Euclidean plane, parametrized by the complex coordinates ${z = v \tau + \mathrm{i} x}$ and ${\bar{z} = v \tau-\mathrm{i} x}$, conformal transformations are implemented as (local) holomorphic and antiholomorphic mappings
\begin{equation}
    z \to w(z), \quad \bar{z} \to \bar{w}(\bar{z}).
\end{equation}

The conformal structure of a given conformally invariant field theory (CFT)~\cite{Ginsparg1988,Senechal1997} is completely determined by its energy-momentum tensor, whose holomorphic and antiholomorphic components are denoted by $T(z)$ and $\bar{T}(\bar{z})$, respectively, together with a special set of local fields $\{\mathcal{O}(z,\bar{z})\}$, known as primary fields.
These primary fields satisfy the following operator product expansions (OPEs) with the components of the energy-momentum tensor:
\begin{equation}
    \begin{split}
        T(z) \mathcal{O}(w,\bar{w}) \sim
            \frac{h \mathcal{O}(w,\bar{w})}{(z-w)^{2}}
            + \frac{\partial_{w} \mathcal{O}(w,\bar{w})}{z-w},
        \\
        \bar{T}(\bar{z}) \mathcal{O}(w,\bar{w}) \sim
            \frac{\bar{h} \mathcal{O}(w,\bar{w})}{(\bar{z}-\bar{w})^{2}}
            + \frac{\partial_{\bar{w}} \mathcal{O}(w,\bar{w})}{\bar{z}-\bar{w}},
    \end{split}
\end{equation}
where the pair of constants $(h,\bar{h})$ are called the conformal dimensions.
With a proper normalization, primary fields satisfy the following OPEs:
\begin{equation}
    \mathcal{O}_{h,\bar{h}}(z,\bar{z})
    \mathcal{O}_{h',\bar{h}'}(w,\bar{w})
        \sim
        \frac{ \delta_{h,h'} \delta_{\bar{h},\bar{h}'} }
        {(z-w)^{2h} (\bar{z}-\bar{w})^{2\bar{h}}}.
\end{equation}

An important characteristic of the given CFT is its central charge $c$, defined from the OPE:
\begin{equation}
\label{eq:T(z)-T(w)-OPE}
    T(z) T(w)
        \sim \frac{1}{2}\frac{c}{(z-w)^{4}}
        + \frac{2 T(w)}{(z-w)^{2}}
        + \frac{\partial_{w}T(w)}{z-w}.
\end{equation}

\subsection{Compact boson CFT}
\label{appendix:Boson-CFT}
The fixed point (Gaussian) action in Eq.~\eqref{eq:Gaussian-Model} of the compact boson field~\cite{Senechal1997,Allen_Senechal_1997,Ginsparg1988,Nersesyan2004}
\begin{equation}
\label{eq:Compact-Bose-Field}
    \phi \sim \phi + 2\pi R,
\end{equation}
where $R = 1 / \sqrt{K}$ is the compactification radius, in a manifestly conformally invariant form, reads:
\begin{equation}
\label{eq:Boson-CFT-Action}
    \mathcal{S}_{\text{B}}^{*}[\phi]
        = \frac{1}{2\pi}
        \int \dd[2]{z}
        \partial_{z} \phi
        \partial_{\bar{z}} \phi,
    \quad
    \dd[2]{z} \equiv v \dd{\tau} \dd{x}.
\end{equation}
The field $\phi(z,\bar{z})$ has a logarithmic correlator:
\begin{equation}
    \langle \phi(z,\bar{z}) \phi(w,\bar{w}) \rangle
        =- \ln[(z-w)(\bar{z}-\bar{w})]
\end{equation}
which suggests a chiral decomposition:
\begin{equation}
    \phi(\tau,x)
        = \varphi(z) + \bar{\varphi}(\bar{z}),
\end{equation}
with
\begin{equation}
    \langle \varphi(z) \varphi(w) \rangle
        =-\ln(z-w),
\end{equation}
and similarly for the antiholomorphic part.
In terms of these, the holomorphic energy-momentum tensor is given by
\begin{equation}
\label{eq:Boson-Energy-Momentum}
    T_{\text{B}}(z) 
        =-\frac{1}{2}
        {:}\partial_{z} \varphi(z)
        \partial_{z} \varphi(z){:},
\end{equation}
which forms the following OPE with itself:
\begin{equation}
\nonumber
    T_{\text{B}}(z) T_{\text{B}}(w)
        \sim \frac{1}{2}\frac{1}{(z-w)^{4}}
        + \frac{2 T_{\text{B}}(w)}{(z-w)^{2}}
        + \frac{\partial_{w}T_{\text{B}}(w)}{z-w}.
\end{equation}
Hence, the central charge of the compact boson $c_{\text{B}} = 1$.

Since the Bose fields have logarithmic correlators, they are not primaries; however, their derivatives are
\begin{equation}
\label{eq:U(1)-Currents}
    J(z) = \mathrm{i}\partial_{z}\varphi(z),
    \quad
    \bar{J}(\bar{z})
        = \mathrm{i}\partial_{\bar{z}}\bar{\varphi}(\bar{z}),
\end{equation}
with conformal dimensions (1,0) and (0,1), respectively.
In fact, $J(z)$ and $\bar{J}(\bar{z})$ are conserved Noether currents corresponding to U(1)$_{\text{L}} \otimes$ U(1)$_{\text{R}}$ symmetry of field shifts:
\begin{equation}
    \varphi \to \varphi + c,
    \quad
    \bar{\varphi} \to \bar{\varphi} + \bar{c},
\end{equation}
where $c,\bar{c}$ are arbitrary constants.

Now, in terms of the chiral components, one defines the dual field
\begin{equation}
    \theta(\tau,x)
        = \varphi(z)-\bar{\varphi}(\bar{z}),
\end{equation}
obeying the duality relations
\begin{equation}
    \partial_{z} \phi = \partial_{z} \theta,
    \quad
    \partial_{\bar{z}} \phi =-\partial_{\bar{z}} \theta.
\end{equation}
The dual field is also compact
\begin{equation}
    \theta \sim \theta + 2\pi \widetilde{R}
\end{equation}
with radius
\begin{equation}
    \widetilde{R} = \frac{2}{R} = \sqrt{4K}.
\end{equation}
The primary operator content of the compact boson, apart from the U(1) currents, is then determined by the following vertex operators:
\begin{equation}
\label{eq:Vertex-Operators-CFT}
    \mathcal{V}_{n,m}(z,\bar{z})
        \equiv
        {:}\exp\!
        \left(
            \mathrm{i} \frac{n}{R} \phi +
            \mathrm{i} \frac{R m}{2} \theta
        \right){:},
    \quad
    n,m \in \mathbb{Z},
\end{equation}
having conformal dimensions:
\begin{equation}
\label{eq:Vertex-Conformal-Dimensions}
    \begin{split}
        h_{n,m}
            & = \frac{1}{2}
            \left( \frac{n}{R} + \frac{Rm}{2} \right)^{2},
        \\
        \bar{h}_{n,m}
            & = \frac{1}{2}
            \left( \frac{n}{R} - \frac{Rm}{2} \right)^{2}.
    \end{split}
\end{equation}
Clearly, currents and the vertex operators satisfy the following OPEs:
\begin{align}
    & T_{\text{B}}(z) J(w)
        \sim \frac{J(w)}{(z-w)^{2}}
        + \frac{\partial_{w}J(w)}{z-w},
    \\
    & T_{\text{B}}(z) \mathcal{V}_{n,m}(z,\bar{w})
    \\
    &\hspace{16mm}
        \sim \frac{h_{n,m} \mathcal{V}_{n,m}(w,\bar{w})}{(z-w)^{2}}
        + \frac{\partial_{w}\mathcal{V}_{n,m}(w,\bar{w})}{z-w},
\end{align}
and similarly for antiholomorphic counterparts.

\subsection{Ising CFT}
\label{appendix:Ising-CFT}
Among the series of unitary minimal models, the Ising CFT represents the simplest example with the central charge $c_{\text{F}} = \flatfrac{1}{2}$.
It is well known~\cite{Senechal1997,Allen_Senechal_1997,Ginsparg1988,Nersesyan2004} to be equivalent to the massless Majorana field theory given by the action in Eq.~\eqref{eq:Ising-CFT-Action}, which in complex coordinates admits the form:
\begin{equation}
    \mathcal{S}_{\text{F}}^{*}[\psi]
        =\frac{1}{2\pi}
        \int \dd[2]{z}
        \left(
            \psi
            \partial_{\bar{z}}
            \psi
            + \bar{\psi}
            \partial_{z}
            \bar{\psi}
        \right).
\end{equation}
The chiral components of the Majorana field $\psi(z)$ and $\bar{\psi}(\bar{z})$ are primaries of conformal dimensions $(\flatfrac{1}{2},0)$ and $(0,\flatfrac{1}{2})$, respectively, satisfying the OPEs:
\begin{equation}
    \psi(z) \psi(w)
    \sim \frac{1}{z-w},
    \quad
    \bar{\psi}(\bar{z}) \bar{\psi}(\bar{w})
    \sim \frac{1}{\bar{z}-\bar{w}},
\end{equation}
while the energy-momentum tensor
\begin{equation}
\label{eq:Fermion-Energy-Momentum}
    T_{\text{F}}(z)
        = - \frac{1}{2} {:} \psi(z) \partial_{z} \psi(z) {:}.
\end{equation}

The mass field, $\varepsilon(z,\bar{z}) \equiv \mathrm{i} \psi(z) \bar{\psi}(\bar{z})$, is also primary of conformal dimensions $(\flatfrac{1}{2}, \flatfrac{1}{2})$.
That is precisely the perturbation in the action in Eq.~\eqref{eq:QI-Action}, which drives the system away from criticality.

The Ising CFT contains two more fields, the order field $\sigma(z,\bar{z})$ and the disorder field $\mu(z,\bar{z})$. They are non-locally related to the fermion fields and both have conformal dimensions $(\flatfrac{1}{16}, \flatfrac{1}{16})$.
In the ordered phase $\langle \sigma \rangle \neq 0$ and $\langle \mu \rangle = 0$, while in the disordered phase $\langle \sigma \rangle = 0$ and $\langle \mu \rangle \neq 0$.
Under the Kramers-Wannier transformation, these two are mapped onto each other $\sigma \leftrightarrow \mu$.
\section{CFTs with additional symmetries}
\label{appendix:CFTs-With-Symmetries}
\subsection{\texorpdfstring{SU(2)$_{k}$}{TEXT} WZNW models}
\label{appendix:SU(2)_k}
Some of the nontrivial generalizations from conventional CFTs are realized by including additional structure in the theory.
CFTs with additional SU(2) invariance, generated by currents ${\vb{J}(z) \equiv (J^{1}(z),J^{2}(z),J^{3}(z))}$ carrying conformal weights $(1,0)$ and satisfying the following SU(2) current algebra at level $k$:
\begin{equation}
\label{eq:SU(2)_k-Current-Algebra}
    J^{a}(z)J^{b}(w)
        \sim
        \frac{\mathrm{i} \varepsilon^{abc} J^{c}(z)}{z-w}
        + \frac{k}{2} \frac{\delta^{ab}}{(z-w)^{2}},
\end{equation}
(and similar expressions for the antiholomorphic sector) serve as the simplest examples of such generalizations.
Then the manifestly SU(2) symmetric holomorphic (Sugawara) energy-momentum tensor admits the form
\begin{equation}
    T(z) = \frac{1}{k+2}
        {:}\vb{J}^{2}(z){:}.
\end{equation}
Such theories, known as SU(2)$_{k}$ WZNW models \cite{Knizhnik1984,Gepner1986,Senechal1997}, are characterized by central charges
\begin{equation}
    c_{\text{SU(2)}_{k}} = \frac{3 k}{k + 2}.
\end{equation}

The simplest case of $k = 1$, having the central charge ${c_{\text{SU(2)}_{1}} = 1}$, can be realized as the compact boson CFT fine tuned to a BKT point.
Indeed, setting $R = \sqrt{2}$ in Eqs.~\eqref{eq:Vertex-Conformal-Dimensions} allows the existence of two additional currents with conformal dimensions $(1,0)$:
\begin{equation}
\label{eq:SU(2)_1-Currents}
    J^{\pm}(z)
        \equiv
        J^{1}(z) \pm \mathrm{i} J^{2}(z)
        = {:}\exp(\pm \mathrm{i} \sqrt{2} \varphi_{\text{L}}(z)){:},
\end{equation}
which together with ${J^{3}(z) \equiv \flatfrac{\mathrm{i} \partial_{z} \varphi(z)}{\sqrt{2}}}$ satisfies the current algebra in Eq.~\eqref{eq:SU(2)_k-Current-Algebra} with $k = 1$.
Thus, the chiral ${\text{U(1)}_{\text{L}}\otimes\text{U(1)}_{\text{R}}}$ symmetry of the Gaussian model is promoted to ${\text{SU(2)}_{\text{L}}\otimes\text{SU(2)}_{\text{R}}}$ symmetry precisely at the BKT point (the full spin-rotational invariant limit in the XXZ language).

Another paradigmatic model corresponds to the case $k = 2$, with the central charge $c_{\text{SU(2)}_{2}} = \flatfrac{3}{2}$, and describes the universality class of three massless Majorana fields, described by an O(3)-symmetric action
\begin{equation}
\label{eq:SU(2)_2-Action}
    \mathcal{S}_{\text{SU(2)}_{2}}[\vb*{\psi}]
        = \frac{1}{2\pi}
        \int \dd[2]{z}
        \left(
            \vb*{\psi} \cdot \partial_{\bar{z}} \vb*{\psi}
            + \vb*{\bar{\psi}} \cdot \partial_{z} \vb*{\bar{\psi}}
        \right).
\end{equation}
In terms of Majorana fields $\vb*{\psi} \equiv (\psi_{1}, \psi_{2}, \psi_{3})$, currents are represented as
\begin{equation}
    J^{a}(z) \equiv - \frac{\mathrm{i}}{2} \varepsilon^{abc} \psi_{b}(z) \psi_{c}(z),
\end{equation}
and similarly for the antiholomorphic sector.
\subsection{Superconformal models}
\label{appendix:SUSY}
In this appendix, we focus on the superconformal field theories as described by the family of actions in Eq.~\eqref{eq:SUSY-Action-Functional}:
\begin{equation}
\label{eq:SUSY-Action-Functional-Appendix}
    \begin{split}
        \mathcal{S}_{R}[\phi,\psi]
            = ~~ & \frac{1}{2\pi}
            \int \dd[2]{z}
            \left(\partial_{z} \phi
            \partial_{\bar{z}} \phi
            +
            \psi \partial_{\bar{z}} \psi + \bar{\psi} \partial_{z} \bar{\psi} \right),
    \end{split}
\end{equation}
parameterized by $R$, or equivalently by the Luttinger parameter $K$~\cite{Bauer2013,Alberton2017,Huijse2015}.

The total energy-momentum tensor
\begin{equation}
    T(z) = T_{\text{B}}(z) + T_{\text{F}}(z)
\end{equation}
(see Eqs.~\eqref{eq:Boson-Energy-Momentum} and~\eqref{eq:Fermion-Energy-Momentum}) satisfies Eq.~\eqref{eq:T(z)-T(w)-OPE} with the total central charge $c_{\text{tot}} = \flatfrac{3}{2}$.

\subsubsection{\texorpdfstring{$\mathcal{N} = (1,1)$}{TEXT} SUSY}
For arbitrary $R \neq \sqrt{2}$, there exists a single super-current, $G(z) \equiv \mathrm{i} \psi(z) \partial \phi(z)$, with conformal weights $(\flatfrac{3}{2},0)$.
It maps fermionic and bosonic states onto each other, in other words, generates supersymmetry, as is seen from its OPEs with the fermion and boson "currents":
\begin{equation}
    \psi(z) G(w) \sim \frac{J(z)}{z - w},
    \quad
    J(z) G(w) \sim \frac{\psi(w)}{(z - w)^{2}},
\end{equation}
where $J(z)$ is the holomorphic U(1) current, Eq.~\eqref{eq:U(1)-Currents}.

Additionally, $G(z)$ satisfies the following OPEs:
\begin{equation}
\label{eq:N=1-Super-Virasoro-Algebra}
    \begin{split}
        T(z) G(z) & \sim
        \frac{3}{2} \frac{G(w)}{(z - w)^{2}}
            + \frac{ \partial_{w} G(w) }{z - w},
        \\
        G(z) G(w) & \sim 
            \frac{2}{3} \frac{c_{\text{tot}}}{(z - w)^{3}}
            + \frac{2 T(w)}{z - w},
    \end{split}
\end{equation}
and similar expressions for antiholomorphic counterparts, which extends the Virasoro algebra in Eq.~\eqref{eq:T(z)-T(w)-OPE} to the ${\mathcal{N} = (1,1)}$ super-Virasoro algebra~\cite{Friedan1984,Bershadsky1985,Mussardo1987,Dixon_Ginsparg_1988} and thus generates the superconformal
symmetry..

In the noninteracting limit $R = 1$, the compactified boson is equivalent to the free massless Dirac field or two massless Majorana fields, as can be easily confirmed using the standard bosonization rules~\cite{Giamarchi2003,Nersesyan2004,Nersesyan2001},
\begin{align}
    \begin{aligned}
        \psi_{1}(z) + \mathrm{i} \psi_{2}(z)
            & \sim
            \sqrt{2}\,
            {:}e^{ -\mathrm{i} \varphi(z) }{:},
        \\
        \bar{\psi}_{1}(\bar{z}) + \mathrm{i} \bar{\psi}_{2}(\bar{z}) 
            & \sim
            \sqrt{2}\,
            {:}e^{ +\mathrm{i} \bar{\varphi}(\bar{z}) }{:}.
    \end{aligned}
\end{align}
Then, denoting $\psi_{3}(z) \equiv \psi(z)$, we find that the superconformal action, Eq.~\eqref{eq:SUSY-Action-Functional-Appendix}, is mapped onto the O(3) symmetric action in Eq.~\eqref{eq:SU(2)_2-Action}:
\begin{equation}
    \mathcal{S}_{R=1}[\phi,\psi]
        = \mathcal{S}_{\text{SU(2)}_{2}}[\vb*{\psi}].
\end{equation}
Therefore, it belongs to the SU(2)$_{2}$ WZNW universality class~\cite{Senechal1997}.
\subsubsection{\texorpdfstring{$\mathcal{N} = (3,3)$}{TEXT} SUSY}
At the BKT point, $R=\sqrt{2}$, the appearance of two additional currents, Eqs.~\eqref{eq:SU(2)_1-Currents}, leads to additional $h = \flatfrac{3}{2}$ super-currents, $G^{1,2}(z) = \sqrt{2}\psi(z) J^{1,2}(z)$, which together with $G^{3}(z) \equiv G(z)$ extend the super-symmetry to the following ${\mathcal{N} = (3,3)}$ algebra~\cite{Schwimmer1987,Ozer2025,Chang1987}:
\begin{equation}
    \begin{split}
        \psi(z)G^{a}(w) & \sim
            \frac{J^{a}(z)}{z - w},
        \\
        J^{a}(z)G^{b}(w) & \sim
            \frac{\delta^{ab}}{\sqrt{2}}
            \frac{\psi(w)}{(z - w)^{2}}
            + \frac{\mathrm{i} \varepsilon^{abc} G^{c}(w)}{z - w},
    \end{split}
\end{equation}
\begin{equation}
    \begin{split}
        T(z)G^{a}(w) & \sim
            \frac{3}{2} \frac{G^{a}(w)}{(z - w)^{2}}
                + \frac{\partial_{w} G^{a}(w)}{z - w},
        \\
        G^{a}(z)G^{b}(w) & \sim
            \frac{2c_{\text{tot}}}{3}
            \frac{\delta^{ab}}{(z - w)^{3}}
            + \frac{2\mathrm{i} \varepsilon^{abc} J^{c}(w)}{(z - w)^{2}}
            \\
            &+
            \frac{
                \delta^{ab} 2T(w)
                + \mathrm{i} \varepsilon^{abc} \partial J^{c}(w)
            }{z - w},
    \end{split}
\end{equation}
and similarly for the antiholomorphic sector.

\bibliographystyle{apsrev4-1}
\bibliography{biblio.bib}

\end{document}